\documentclass[aps,prd,10pt,twocolumn,superscriptaddress,nofootinbib,showkeys,showpacs,altaffilletter,floatfix]{revtex4-1}

\usepackage{graphicx}
\usepackage{dcolumn}
\usepackage{amssymb}
\usepackage{amsmath}
\usepackage{amsfonts}
\usepackage{amsbsy}
\usepackage{color}
\usepackage{rotating}
\usepackage[english]{babel}
\usepackage{soul}
\usepackage{multirow}
\usepackage{comment}
\usepackage{overpic}
\usepackage{ifxetex}
\ifxetex
   \usepackage{polyglossia}
   \setdefaultlanguage{english}
   \usepackage{fontspec}
\else
   \usepackage[english]{babel}
   \usepackage[utf8]{inputenc}
\fi

\usepackage{cancel}
\usepackage{verbatim} 

\newcommand{\be}{\begin{equation}}
\newcommand{\ee}{\end{equation}}
\newcommand{\bea}{\begin{eqnarray}}
\newcommand{\eea}{\end{eqnarray}}

\newcommand*{\rom}[1]{\expandafter\@slowromancap\romannumeral #1@}
\makeatother

\begin{document}

\title{Observational constraints on cosmological solutions of $f(Q)$ theories}

\author{Ismael Ayuso} 
\email{iayuso@fc.ul.pt}
\affiliation{Departamento de Física and Instituto de Astrofísica e Ciências do Espaço, Faculdade de Ciências, Universidade de Lisboa, Edifício C8, Campo Grande, 1769-016 Lisboa, Portugal}
\author{Ruth Lazkoz}
\email{ruth.lazkoz@ehu.es}
\affiliation{Department of Theoretical Physics, University of the Basque Country UPV/EHU, P.O. Box 644, 48080 Bilbao, Spain}
\author{Vincenzo Salzano}
\email{vincenzo.salzano@usz.edu.pl}
\affiliation{Institute of Physics, Faculty of Mathematics and Physics, University of Szczecin, Wielkopolska 15, 70-451 Szczecin, Poland}

\date{\today}

\begin{abstract}
Over the last years some interest has been gathered by $f(Q)$ theories, which are new candidates to replace Einstein's prescription for gravity. The non-metricity tensor  $Q$ allows to put forward the assumption of a free torsionless connection and, consequently, new degrees of freedom in the action are taken into account. This work focuses on a class of $f(Q)$ theories, characterized by the presence of a general power-law term which adds up to the standard  (linear in) $Q$ term in the action, and on new cosmological scenarios arising from them. Using the Markov chain Montecarlo method we carry out statistical tests relying upon background data such as Type Ia Supernovae luminosities and direct Hubble data (from cosmic clocks), along with Cosmic Microwave Background shift and Baryon Acoustic Oscillations data. This allows us to perform a multifaceted comparison between  these new cosmologies and the (concordance) $\Lambda$CDM setup. We conclude that,  at the current precision level, the best fits of our $f(Q)$ models correspond  to values of their specific  parameters which make them hardly distinguishable from our General Relativity ``échantillon", that is $\Lambda$CDM.
\end{abstract}

\maketitle

\maketitle

\section{Introduction}

The Universe is dominated by the weakest force of all, that is, the gravitational interaction,  and it is through a deep comprehension of the latter  that we will  improve our understanding of the former. Currently, the most successful theory concerning the behaviour and evolution of the Universe is General Relativity (GR) \cite{Einstein:1915ca,Einstein:1915by}, which passes  all tests up to the solar system scale  with flying colors \cite{Will:2014kxa,Berti:2015itd}. However, when we try to explain the physics of the cosmos on larger scales, it becomes mandatory to  consider more exotic kinds of energy and matter than the standard sources of geometry. In this way, and under the uncontroversial assumption of  homogeneity  and isotropy in a GR ruled Universe, the $\Lambda$CDM model provides a quite worthy chronicle of our Universe driven by new unusual components (as we have anticipated): dark matter and dark energy. The first one accounts proficiently for structure formation and evolution, whereas, the second one performs quite well at explaining the current accelerated expansion of the Universe.

Nevertheless, we cannot describe the physics of our Universe by just scraping its surface: when we examine the $\Lambda$CDM model in the light of the whole assortment of evidences available, we realize that, unfortunately, its two main components lead to new problems \cite{Carroll:2000fy, Peebles:2002gy, Weinberg:1988cp, Padmanabhan:2002ji, Carroll:2003qq, Bianchi:2010uw, Bull:2015stt}. In particular, volumes have been written about the shortcomings of the specific type of dark energy on which the model relies, the cosmological constant ($\Lambda$). Needless to say it is not our purpose to dwell into this matter, but rather adopt a different perspective.

Modified theories of gravity have precisely been proposed along the years to tackle those and other challenges \cite{Capozziello:2002rd,Clifton:2011jh,Joyce:2014kja,Harko:2018ayt}.
Among the different starting points considered in the vast literature,  we will embrace that of metric-affine geometry, which generalizes the Riemannian geometry approach adopted in GR. The connection then becomes a non-standard free variable at the same level as the metric, and hence it is not necessarily of the Levi-Civita type.
Broadly speaking, this freedom in the features of the connection  brings a rich phenomenology related to the transformations which objects of physico-mathematical nature undergo along a displacement \cite{BeltranJimenez:2019tjy,Jimenez:2019ghw}.

Specifically, we are going to study theories of modified gravity based on non-metricity \cite{Mol:2014ooa}, a quantity which measures how the length of a vector changes when it is transported.  After offering  new views on some theoretical aspects of this ample framework, we will resort to observational tests so as to draw conclusions on the (statistical and physical) reliability of some specific models.

The connection between theory and observations arises from the  possibility to write useful prescriptions for cosmological applications. Specifically, just following pertinent generalizations of otherwise standard procedures,  one can compute the Hubble parameter analytically \cite{Jimenez:2019ovq}  for a class of non-metricity
spacetime geometries. The statistical analysis is performed using  state-of-the-art cosmological probes under the assumption of some specific matter-energy content  so that the candidate models offer (a priori) viable modified gravity candidates to describe the cosmological background. The procedure progresses from the initial expression of the Hubble parameter to the construction of various cosmological distances (luminosity distance, angular diameter distance, etc.). Interestingly, the GR limit  of these modified gravity frameworks will be easily recognisable, thus allowing for a neat statistical analysis.

Let us now offer a guide to the organization of this manuscript. In section \ref{STG} we will produce a convenient introduction to non-metricity theories of modified gravity. In this very section, and for the sake of motivation of further steps, we will revisit how one can construct genuine settings based solely on non-metricity   which, however, do mimic GR perfectly at all levels.

We will move then to section \ref{Cosmolofiesfq} and build upon the previous discussion to offer more general scenarios, still based on  non-metricity, but not necessarily fully equivalent to GR. Upon very general hypotheses we will be able to present the new versions of the Friedman and Raychaudhuri equations. An intermediate step within this section deals with possible equivalences between these new frameworks and GR at the cosmological level. This gives us a flavour of how peculiar the consequences of non-metricity are. The findings of this exercise will serve as an inspiration to propose (obtain) the Hubble parameter of the final specific model  we shall work with. Our models will display a
characteristic set of parameters: all but one are very standard, and it is upon switching  off the unusual one that we recover the customary GR sort of cosmological evolution.

A statistical analysis based on the Markov Chain Monte Carlo (MCMC) procedure (see section \ref{parameters}) will yield tight enough constraints on the parameters, and results will be shown in section \ref{Results}. The best fits, errors and statistical criteria will be driven by supernovae type Ia luminosity data and modulated by direct Hubble (cosmic clocks), CMB shift and BAO data (head to section \ref{data} for further details). Lastly, we will dissertate about our conclusions in section \ref{Conclusions}.


\section{Symmetric Teleparallel Gravity}
\label{STG}

Currently, the gravitational interaction is interpreted as a geometrisation of a manifold allowed by the equivalence principle. Put in other words, gravity is a physical  manifestation of the mathematical conformation of that manifold. For the case of GR, a spacetime comes about upon endowing a manifold with a prescription to measure distances that mathematically takes the form of the metric tensor $g_{\mu\nu}$ and is the only object where gravitational effects are encoded.

Interestingly, though, the metric does not happen to be the only fundamental object allowing to characterize the  geometry of a manifold. Furthermore, there is an affine structure  associated with how objects move about the manifold; this is represented by a connection, $\Gamma^{\alpha}_{\mu\nu}$, a degree of freedom (dof) which may be subject to certain specifications imposed by the theoretical setting. For instance, in GR, the connection amounts to a combination of (conventional) derivatives of the metric tensor (Christoffel symbols), but other options are possible if one ventures beyond the classical realm in gravity \cite{Hehl:1994ue}.

Hence, provided that the connection is not (so to speak) geometrically trivial, one can define two further fundamental objects  conveying additional relevant information about it \cite{Jarv:2018bgs}. The first one is  the non-metricity tensor:
\begin{equation}
    Q_{\alpha\mu\nu}\equiv\nabla_\alpha g_{\mu\nu}.
\end{equation}
The second one is the torsion, which stems from the antisymmetric part of the connection:
\begin{equation}
    T^{\alpha}_{\mu\nu}\equiv\Gamma^\alpha_{\mu\nu}-\Gamma^\alpha_{\nu\mu}. 
\end{equation}
If the connection is symmetric and metricity holds (i.e. when the torsion and non-metricity tensors vanish), we recover the Levi-Civita connection, thus returning to gravity ``à la Einstein" (in which the metric is the only dof). Conversely, the metric and the connection could be considered as independent objects whose relations would be given by the field equations. This is the so-called Palatini formalism \cite{Capozziello:2007ec, Olmo:2011uz,Olmo:2005hc, Sotiriou:2006qn, Harko:2018ayt}.

In addition, and in view of the previous discernments, a general connection $\Gamma^{\alpha}_{\mu\nu}$ can be decomposed as follows \cite{Jarv:2018bgs, BeltranJimenez:2019tjy}:
\bea
\Gamma^{\alpha}_{\mu\nu}=\left\{\substack{\alpha\\ \mu\nu}\right\}+K^{\alpha}_{\;\;\mu\nu}+L^{\alpha}_{\;\;\mu\nu},
\eea
where
\bea
K^{\alpha}_{\;\;\mu\nu}&=&\frac{1}{2}T^{\alpha}_{\;\;\mu\nu}+T^{\;\;\;\alpha}_{(\mu\;\;\;\nu)}
\eea
is the contortion, and
\bea
L^{\alpha}_{\;\;\mu\nu}&=&\frac{1}{2}Q^{\alpha}_{\;\;\mu\nu}-Q^{\;\;\;\alpha}_{(\mu\;\;\;\nu)}
\eea
is the  disformation. The latter will be the relevant magnitude for this work, as it measures how much the symmetric part of the (general) connection deviates from the Levi-Civita connection, $
\left\{\substack{\alpha\\ \mu\nu}\right\}
$. In any case, and as usual, the curvature will be determined by the Riemann tensor in this way:
\bea
R^{\alpha}_{\;\;\beta\mu\nu}(\Gamma)=\partial_\mu\Gamma^{\alpha}_{\;\;\nu\beta}-\partial_\nu\Gamma^{\alpha}_{\;\;\mu\beta}+\Gamma^{\alpha}_{\;\;\mu\lambda}\Gamma^{\lambda}_{\;\;\nu\beta}-\Gamma^{\alpha}_{\;\;\nu\lambda}\Gamma^{\lambda}_{\;\;\mu\beta}.\nonumber\\\label{Riemann}
\eea

However, all this information will be of little use  without a theory of gravity, so let us make some progress by introducing GR through the Hilbert-Einstein action:
\bea
S_{GR}=\frac{1}{16\pi G}\int{d^x\sqrt{-g}R(\{\})},
\label{GR}
\eea
where $R(\{\})$ is the scalar of curvature upon the specific choice of the Levi-Civita connection in Eq.~\eqref{Riemann}. This action has a special property. Let us think for a moment that we readdress it but  using rather now  the general scalar of curvature $R(\Gamma)$, instead of $R(\{\})$, and then study this ``alternative action" in the Palatini formalism.
Upon the exercise of varying the action with respect to the metric on the one hand, and the connection on the other hand, we obtain two sets of equations which are not blind to each other: they specifically fix the connection to be the Levi-Civita one!
Summarizing, if  GR is our  gravitational theory, both formalisms  yield the same equations of motion because non-metricity and torsion vanish. Nevertheless, this is but an exception of GR which will not be true for theories of modified gravity in general, and yet it can be used to our advantage. Indeed,  incorporating the connection as a new dof on the framework of modified gravity opens a complete new range of theoretical settings to explore.

The next stop in our journey is to study how the Riemmann tensor is transformed by a shift of the connection of the form $\hat\Gamma^{\alpha}_{\;\;\beta\mu\nu}=\Gamma^{\alpha}_{\;\;\mu\nu}+\Omega^{\alpha}_{\;\;\mu\nu}$ where $\Omega^{\alpha}_{\;\;\mu\nu}$ is an arbitrary tensor which encodes the transformation. Specifically
\bea
\hat{R}^{\alpha}_{\;\;\beta\mu\nu}(\hat\Gamma)=&&R^{\alpha}_{\;\;\beta\mu\nu}(\Gamma)+T^{\lambda}_{\;\;\mu\nu}\Omega^{\alpha}_{\;\;\lambda\beta}+\nonumber\\
&&2\nabla_{[\mu}\Omega^{\alpha}_{\;\;\nu]\beta}+2\Omega^{\alpha}_{\;\;[\mu\mid\lambda\mid}\Omega^{\lambda}_{\;\;\nu]\beta}.
\label{transformationR}
\eea
where $\nabla$ is the covariant derivative associated to $\Gamma$.

Upon  inspection of the latter, one can follow on  the steps of \cite{BeltranJimenez:2019tjy}
to end up realising the existence of a theory which is fully equivalent to GR (as obtained from the Levi-Civita connection) but coming instead and solely from the  disformation (the part of the connection related to the non-metricity)\footnote{Note that in this theory the torsion plays no role whatsoever.}.
This requirement translates into  $\Gamma=\{\}$, $T^{\lambda}_{\;\;\mu\nu}=0$ and $\Omega^{\alpha}_{\;\;\mu\nu}=L^{\alpha}_{\;\;\mu\nu}$. If we replace those expressions into Eq.~\eqref{transformationR} it becomes
\bea
\!\!\!\!\!\!\!\!\!\!\!\!{R}^{\alpha}_{\;\;\beta\mu\nu}(\hat\Gamma)=&&R^{\alpha}_{\;\;\beta\mu\nu}(\left\{\right\})+2\nabla^{\{\}}_{[\mu}L^{\alpha}_{\;\;\nu]\beta}+2L^{\alpha}_{\;\;[\mu\mid\lambda\mid}L^{\lambda}_{\;\;\nu]\beta},
\label{transformationRL}
\eea
where the super-index in $\nabla^{\{\}}$ specifies the covariant derivative defined from the Levi-Civita connection. The latter and other intermediate steps conform our alternative and hopefully pedagogical explanation about how to  build such theory.

Again, inspection suggests additional requirements are needed in order to eventually reach the goal, so in this spirit let us demand our spacetime to be  flat  by the constraint  $R^{\alpha}_{\;\;\beta\mu\nu}(\hat\Gamma)=0$, which reduces the connection to the Weitzenböck form \cite{Weitzenbock,Aldrovandi:2013wha}. The theories formulated in these frameworks are referred to as teleparallel due to a well-defined notion of parallelism\footnote{As a bonus, that this formulation can be regarded as a “translational gauge theory” \cite{ Maluf:1994ji, Maluf:1995re, deAndrade:1997gka, Muench:1998ay}.} as a consequence of the vanishing of total curvature  \cite{Hayashi:1979qx,Mielke:1992te,BeltranJimenez:2019tjy}. 


As the literature proves, in this symmetric teleparallel  framework \cite{Nester:1998mp}, it is possible to find a theory which reproduces exactly GR but through the non-metricity  tensor, $Q_{\alpha\mu\nu}$ which mediates the gravitational force, instead of resorting to curvature for that task \cite{BeltranJimenez:2019tjy}. This special case is called Symmetric Teleparallel Equivalent General Relativity (STEGR).

But we wish to throw further light into  this fact, and to that end we perform a contraction of Eq.~\eqref{transformationRL}  which can be used to  rewrite $R(\{\})$  (the piece we eventually need to show the equivalence at the Lagrangian level):
\bea
0=R(\{\})&+&\nabla^{\{\}}_\alpha\left(Q^\alpha-\tilde{Q}^\alpha\right)+\frac{1}{4}Q^{\alpha\beta\gamma}Q_{\alpha\beta\gamma}\nonumber\\
&-&\frac{1}{2}Q^{\gamma\alpha\beta}Q_{\alpha\gamma\beta}
-\frac{1}{4}Q^\alpha Q_\alpha+\frac{1}{2}\tilde{Q}^\alpha Q_\alpha,
\label{0RQ}
\eea
where $Q_{\alpha}=Q_{\alpha\;\;\;\mu}^{\;\;\;\mu}$ and $\tilde{Q}_\alpha=Q^{\mu}_{\;\;\;\alpha\mu}$. It is convenient now to simplify the latter expression by defining the non-metricity scalar $Q$ as\footnote{However, the most general even-parity second order quadratic form of the non-metricity is:
\bea
\mathcal{Q}=\frac{c_1}{4}Q_{\alpha\mu\nu}Q^{\alpha\mu\nu}-\frac{c_2}{2}Q_{\alpha\mu\nu}Q^{\mu\alpha\nu}-\frac{c_3}{4}Q_\alpha Q^\alpha\nonumber\\
+\frac{c_4}{2}Q_\alpha \tilde{Q}^\alpha+(c_5-1)\tilde{Q}_\alpha \tilde{Q}^\alpha,\nonumber
\eea
which is a generalisation of Eq.~\eqref{Q} that gets recovered by setting  $c_1=c_2=c_3=c_4=1$.}
:
\bea
\!Q=\frac{1}{4}Q^{\alpha\beta\gamma}Q_{\alpha\beta\gamma}-\frac{1}{2}Q^{\gamma\alpha\beta}Q_{\alpha\gamma\beta}-\frac{1}{4}Q^\alpha Q_\alpha+\frac{1}{2}\tilde{Q}^\alpha Q_\alpha.\nonumber\\
\label{Q}
\eea

Therefore, for  symmetric (torsionless) and flat constraints we get:
\bea
R({\{\}})=-Q-\nabla^{\{\}}_\alpha(Q^\alpha-\tilde Q^\alpha),
\label{RQ}
\eea
where $\nabla^{\{\}}_\alpha$ is a total divergence term. Continuing with this lengthy scheme we can verify that, if we take the scalar of curvature obtained from the Levi-Civita connection appearing in Eq.~\eqref{GR} and we replace it with Eq.~\eqref{RQ}, we in fact build a theory which is equivalent to GR up to a total derivative in the action, which, in any case, does not contribute to the equations of motion:
\bea
S_{GR}&=&\frac{1}{16\pi G}\int{d^4 x\sqrt{-g}R({\{\}}})=\nonumber\\
&-&\frac{1}{16\pi G}\int{d^4 x\sqrt{-g}Q}=S_{\rm STEGR},
\eea
Therefore, the STEGR theory and GR are equivalent frameworks of gravity but formulated  with $R$ and $Q$ respectively. Obviously, cosmological models following from these two settings are completely identical, and cosmological observations  would no offer hints as to which is the underlying theory.  However, we now are in a position which allows us to go one step beyond and build $f(Q)$ models, close enough to GR to make sense, but at the same time, different enough so that small modifications could ideally be spotted.  In this sense we want to stress that theories stemming from actions based on $Q$ have only been explored superficially in what regards observations; in fact, no cosmological tests can be found in the literature. For this reason, we feel that statistical examinations of the free parameters of $f(Q)$ cosmologies in the light of astrophysical data could help us discuss the suitability of such scenarios.


\section{$f(Q)$ cosmologies}
\label{Cosmolofiesfq}

Following  the justification offered in the previous sections, we generalize now the STEGR action as follows:
\begin{equation}
    S=\int{d^4 x\sqrt{-g}\left[-\frac{1}{2}f(Q)+\mathcal{L}_{M}\right]},
\end{equation}
where STEGR is directly recovered for $f(Q)=Q/8\pi G$.

Analogously to \cite{Jimenez:2019ovq}, we are going to work in the coincident gauge, which allows to use a null connection, i.e.
\bea
{_{cg}\Gamma}^\alpha_{\mu\nu}=0 \;\;\; \rightarrow \;\;\; \nabla_\alpha g_{\mu\nu}=\partial_\alpha g_{\mu\nu}.
\eea
Besides, we will consider a spatially flat FLRW spacetime, as the most used  homogeneous an isotropic standard spacetime to describe the Universe on large scales,
\bea
ds^2=-N^2(t)dt^2+a^2(t)\left[dx^2+dy^2+dz^2\right]
\eea
for which, the non-metricity scalar reads
\bea
Q=6\frac{H^2}{N^2}.
\label{non-metricity scalar}
\eea
As shown in \cite{BeltranJimenez:2018vdo}, $f(Q)$ theories let us fix a particular lapse function because $Q$ retains a residual time-reparameterization invariance, in spite of which we have already used diffeomorphisms to select the coincident gauge. Therefore, and for simplicity, we will take this to our advantage and choose $N(t)=1$. Then, the cosmological equations become
\bea
6 f_Q H^2-\frac{1}{2}f&=&\rho, \label{Friedmann}\\
(12 H^2 f_{QQ}+f_Q)\dot H&=&-\frac{1}{2}(\rho+p) \label{Ray},
\eea
where the sub-index denotes derivatives with respect to $Q$. Examination shows that we can reproduce exactly the GR background behaviour under the prescription
\bea
Q f_Q-\frac{1}{2}f=\frac{3 H^2}{8\pi G}=\frac{Q}{16\pi G}\,,
\eea
which implies
\bea
f(Q)=\frac{1}{8\pi G}\left(Q+M\sqrt{Q}\right)
\label{fqgr},
\eea
where $M$ is a constant which can be interpreted as a mass scale \cite{Jimenez:2019ovq}. The analogy between the particular case with $M=0$ and GR should be not surprising at all, because it corresponds to the STEGR framework we already discussed in the previous section. But the $M\not =0$ case represents a whole class of theories with the same background as GR whose differences do not show up at the background level, although they do it at the perturbation level.

An alternative route is to put forward an ansatz for $f(Q)$ that includes Eq.
\eqref{fqgr} as a particular case in the hope that this analytical extension  will let us integrate Eq.~\eqref{Friedmann} and Eq.~\eqref{Ray} and so progress will be made possible. This happens to be the case for the  proposal presented in \cite{Jimenez:2019ovq}:
\bea
f(Q)=\frac{1}{8\pi G}\left[Q-6\lambda M^2\left(\frac{Q}{6M^2}\right)^\alpha\right]
\label{f},
\eea
where $\lambda$ and $\alpha$ are dimensionless parameters and $M\not = 0$.  Interestingly, different values of  $\alpha$ can be chosen to construct solutions depicting new early or late universe behaviours.

From Eq.~\eqref{f} and using Eq.~\eqref{non-metricity scalar}, the Friedmann equation Eq.~\eqref{Friedmann} can be integrated to yield
\bea
H^2\left[1+(1-2\alpha)\lambda\left(\frac{H^2}{M^2}\right)^{\alpha-1}\right]=\frac{8\pi G}{3}\rho.
\eea
Note that a background evolution identical to that of GR is recovered for either $\lambda=0$ or   $\alpha=1/2$. In addition, the case for $\alpha=1$ follows the same dynamics than GR after a  redefinition of $G$.

In this paper, though,  we will follow \cite{Jimenez:2019ovq} and set the focus on the  $\alpha=-1$ case, which leads to a solution with two possible branches:
\bea
H^2_{\pm}=\frac{4\pi G}{3}\rho\left(1\pm\sqrt{1-\frac{27\lambda M^4}{(4\pi G \rho)^2}}\right), \label{hubble}
\eea
where $\rho$ is the sum of all energy densities (as customary it will be regarded as positive). The  correction with respect to the GR case becomes larger as $\rho$ decreases, so the new degree of freedom plays the effective role of  dark energy. Note the considerable level of non-linearity at play in Eq.~\eqref{hubble}.

From now on, and in order to be able to exploit the predicting capabilities of an assortment of cosmological data sets, we are going to consider our Universe's evolution is driven by the three usual kinds of matter-energy: cosmic dust, radiation and a cosmological constant.

The explicit presence of a cosmological constant might seem redundant or unnecessary, as we rather want to explore geometric corrections that mimic its effect. However, it will become clear that, as far as statistical comparisons are concerned, the presence of such a term renders the whole analysis far more palpable.

Additionally, the whole lot of standard matter-energy fields will satisfy the continuity equation, i.e.
\bea
\dot{\rho}=-3H(\rho+p),
\label{continuityequation}
\eea
where the dot denotes derivation with respect to cosmic time. With our choices  $\rho$ reads
\bea
\rho=\frac{3H_0^2}{8\pi G}\left[\Omega_\Lambda+\Omega_m(1+z)^3+
\Omega_r(1+z)^4\right], \label{rhoeq}
\eea
where we let $\Omega_\Lambda$, $\Omega_m$ and  $\Omega_r$ stand for current values of the fractional densities of the three cosmological fluids comprising our matter-energy lot.
Under this whole set of prescriptions the way to recover the $\Lambda$CDM setting, for the positive branch, is
\bea
H^2_{+}\vert_{\Omega_Q=0}=H^2_{\Lambda{\rm CDM}}.
\eea

We will also resort to the customary  ``normalization" of  the Hubble parameter so as to lighten the notation whenever possible:
\bea
E(z=0)=\frac{H_{\pm}(z=0)}{H_0}=1.
\eea
This gives us
\bea
-\frac{M^4\lambda}{H_0^4}=\frac{1}{3}\left(1-\Omega_\Lambda-\Omega_m-\Omega_r\right)\equiv\frac{\Omega_Q}{3}, \label{omegaqdef}
\eea
which can be seen to hold for both branches (obviously, the standard $\Lambda$CDM normalization condition follows from choosing $\Omega_Q$). Besides, and consequently, we will say that $\Omega_\Lambda,\Omega_m$ and $\Omega_r$ are primary parameters, in contrast to $\Omega_Q$ which is a derived one. 

In addition, let us remark a peculiarity from the previous normalization. When it is carried out for the positive branch, one of the intermediate steps is:
\bea
\sqrt{1-\frac{12\lambda M^4}{H_0^4(\Omega_\Lambda+\Omega_m+\Omega_r)^2}}=\frac{2-\Omega_\Lambda-\Omega_m-\Omega_r}{\Omega_\Lambda+\Omega_m+\Omega_r}.\nonumber\\
\eea
Because the left side of this equation is positive, the right side must be as well. Therefore, it is also necessary to impose the condition $0<\Omega_\Lambda+\Omega_m+\Omega_r<2$, and this condition lets $\Omega_Q$ take negative values.  By contrast, a healthy behaviour enabling the normalization  in the negative branch would demand either $\Omega_\Lambda+\Omega_m+\Omega_r>2$ or $\Omega_\Lambda+\Omega_m+\Omega_r<0$   (clearly the second condition makes no sense physically).

After these convenient remarks we can use Eq.~\eqref{omegaqdef} to rewrite the Hubble parameter as a function of the free parameters  to be fitted at a later stage in this work:
\bea
&&H_{\pm}^2=\frac{H_0^2}{2}\left[\Omega_\Lambda+\Omega_m(1+z)^3+\Omega_r(1+z)^4\right]\nonumber\\
&&\times\left(1\pm\sqrt{1+\frac{4\Omega_Q}{\left[\Omega_\Lambda+\Omega_m(1+z)^3+\Omega_r(1+z)^4\right]^2}}\right).~
\label{Hfinal}
\eea

Up to this point we have presented a discussion as general as possible, but in the reminder we will consider the positive branch only, because the  negative one depicts a Hubble parameter which decreases as $z$ increases and therefore seems quite unlikely to match observational evidences.

Obviously, at high redshifts the contribution of $\Omega_Q$ becomes negligible and one recovers the usual $\Lambda$CDM Hubble parameter, as we have already mentioned. However,  at the asymptotic future one rather has
\bea
\underset{\rho_m,\rho_r\rightarrow 0}{\textit{lim}}H_+^2=\frac{1}{2}H_0^2\left[\Omega_\Lambda+\sqrt{\Omega_\Lambda^2+4\Omega_Q}\right],
\label{Hds}
\eea
and for the extremal case without a cosmological constant, i.e. $\Omega_\Lambda=0$, $H_{ds}^2=H_0^2\sqrt{\Omega_Q}$.

It is clear that in order to guarantee the physicality of our expression in this particular regime we should impose more restrictive conditions, as the positivity of $\Omega_Q$. Having made this remark, we will rather let  data speak for themselves and see whether observations end up favouring parameter values which will cause trouble in the negative redshift region (of which they really offer no control).

From a wider perspective, one could also wonder about the particular case of our $f(Q)$ model that follows from setting $\Omega_\Lambda=0$, thus letting the effects of non-metricity account entirely for the dark energy sector allowing us to waive the presence of a cosmological constant or any other form of dark energy altogether. Although this scenario might seem too optimistic, we will examine it too.

\section{MCMC analysis} \label{parameters}

One of the main objectives of this paper is to obtain as tight as possible constraints on the parameters of the $f(Q)$ model under study. Results  will throw some light as to whether non-metricity effects are compatible with observations, thus opening a new window of interest on the possibility
of an underlying modified gravity description of
our universe. In order to narrow down our conclusions we will combine different background astrophysical probes of known statistical relevance.

As it has been already stressed, we are sticking to the positive branch Eq.~\eqref{Hfinal} of these new cosmological scenarios.
The tests will be implemented using
an MCMC code \cite{Lazkoz:2010gz,Capozziello:2011tj},
which, upon minimization of a total $\chi^2$, will produce proficient fits of the values of $\Omega_m,$ $h,$ $\Omega_\Lambda,$ $\Omega_Q$ and $\Omega_b$; and, by the same token, 
this analysis will also produce  selection criteria permitting us to draw unimpaired conclusions.

\subsection{Priors}

Our results will be obtained under the assumption of some priors, which  give some room to modified gravity features, enforce the choice of the right branch and preclude nonphysical behaviour and  pronounced departures from the well established standard evolution (the $\Lambda$CDM golden pattern). Specifically
we assume a uniform uninformative probability for values of the parameters within the intervals defined by:
\begin{itemize}
\item $0<\Omega_r<\Omega_b<\Omega_m$
\item $0<h<1$
\item $0<\Omega_m+\Omega_\Lambda+\Omega_r<2$
\end{itemize}
We stress again that according to our earlier discussion, and in view of these priors, $\Omega_Q$ can be either positive or negative. 

But the best fits and errors alone are not all the information we can infer from the MCMC procedure. Indeed, we can also learn about the reliability of the model by invoking the evidence $\mathcal{E}$, also dubbed   marginal likelihood or integrated likelihood. It is defined as follows:
\bea
\mathcal{E}=P(D |\mathcal{M})=\int P(D |\theta,\mathcal{M})P(\theta | \mathcal{M})d\theta\, ,
\eea
i.e. it estimates the support of the (measured) data $D$ for a given model $\mathcal{M}$ once all possible values for the parameters $\theta$ have been considered. The evidence is generally recognized as the most reliable statistical tool for model comparison in cosmology \cite{Nesseris:2012cq}, provided that wide-enough priors are chosen.

Specifically, the performance of different models is compared through the Bayes factor, that is, the ratio of their evidences:
\bea
\mathcal{B}^{i}_{j}=\frac{\mathcal{E}_i}{\mathcal{E}_j}.
\eea
 In gross terms if $\mathcal{B}^{i}_{j}>1$, i.e ${\mathcal{E}_i}>{\mathcal{E}_j}$ for the measured data $D$, then model $\mathcal{M}_i$ is preferred over  model $\mathcal{M}_j$.

However, it is difficult to quantify how much better (or worse) is one scenario as compared to the other. Jeffreys' Scale \cite{Jeffreys61} is typically adduced in this regard. According to that criterion, if $\ln \mathcal{B}^{i}_{j}<1$, the evidence in favor of the model $\mathcal{M}_i$ is not significant; if $1<\ln \mathcal{B}^{i}_{j}<2.5$, the evidence is substantial; if $2.5<\ln \mathcal{B}^{i}_{j}<5$, it is strong; and if $\ln \mathcal{B}^{i}_{j}>5$, it is decisive. Nevertheless, Jeffreys' scale is not completely flawless, as discussed in \cite{Nesseris:2012cq}.

\subsection{Cosmographic parameters}

On the other hand, it is customary to examine other quantities which offer a clearer picture of the
evolutionary features of the particular FLRW spacetime under study. In fact, once constraints on $\Omega_m,$ $h,$ $\Omega_\Lambda,$ $\Omega_Q$ and $ \Omega_b$ are obtained through our MCMC procedure, we can also draw inferences on the well-known  cosmographic parameters, which follow from the Taylor expansion of the scale factor:
\bea
a(t)=&&a_0\left[1+H_0 \Delta t-\frac{q_0}{2}H_0^2 \Delta t^2+\frac{j_0}{3!}h_0^3\Delta t^3\right.\nonumber\\ &&\left.+\frac{s_0}{4!}h_0^4\Delta t^4+O\left(\Delta t^5\right)\right].
\eea
In the latter we have defined $\Delta t=t-t_0$ while $q_0$, $j_0$ and $s_0$ are the so-called  deceleration, jerk and snap parameters respectively evaluated at $t_0$ (present time)
\cite{Visser:2004bf,Cattoen:2007sk}. Explicit expressions to evaluate them can be found in many references and we just reproduce them here for the benefit of the readers:
\bea
&q(t)=-\displaystyle\frac{a\ddot{a}}{\dot{a}^2} \;\rightarrow \; &q(z)=-1+(1+z)\frac{E'(z)}{E(z)}\, ,\\
&j(t)=\displaystyle\frac{a^2\dot{\ddot{a}}}{\dot{a}^3} \; \rightarrow \; &j(z)=(1+z)^2\frac{E''(z)}{E(z)}+q^2(z)\, ,\qquad\\
&s(t)=\displaystyle\frac{a^3\ddot{\ddot{a}}}{\dot{a}^4} \; \rightarrow \;& s(z)=-(1+z)j'(z)
-2j(z)-3q(z)j(z)\, ,\nonumber\\
\eea
where the dot and the prime denote differentiation with respect to cosmic time and   $z$ respectively.

\section{Observational data and statistical analysis}\label{data}



In this section, we put forward the cosmological data used in this work for an observational scrutiny of both $\Lambda$CDM and the $f(Q)$ given by Eq.~\eqref{Hfinal}. Specifically, we use Type Ia Supernovae with Pantheon data, the expansion rate data from early-type galaxies as cosmic chronometers with \textit{Hubble} data, Cosmic Microwave Background shift parameters from \textit{Planck} 2018, and Baryon Acoustic Oscillations data to this purpose.

We are going to proceed as follows. First, we will provide details on how to apply each probe on its own, that is, we will explain how we will construct each separate $\chi^2$  contributing to the  total one (the sum of all previous ones). Then we will find the values of the parameters which minimize each of those individual contributions (with the pertinent errors) in order to appreciate the contribution of each data set, and after this we will repeat the procedure but using the total $\chi^2$.

Our best fits report will be arranged in a table so as to make the conclusions readier to be drawn. We will also provide the complementary  visual support of confidence contours, which inform us pictorially on the degree of correlations among parameters, the tightness of the constraints each data set suggests, and various other interpretation tools.

\subsection{Pantheon Supernovae data}

This sample is one of the latest Type Ia Supernovae (SNeIa) compilations \cite{Scolnic:2017caz} and  it contains 1048 SNeIa at redshift $0.01<z<2.26$. The constraining power of SNeIa becomes manifest when  used as standard candles. This can be implemented through the use of the distance modulus:
\bea
\mathcal{F}(z,\mathbf{x})_\text{theo}=5\log_{10}\left[D_L(z,\mathbf{x})\right]+\mu_0,
\eea
where $D_L$ is the luminosity distance given by:
\bea
D_L(\mathbf{x})=(1+z)\int_0^z{\frac{c\, dz'}{H_0 E(z',\mathbf{x})}},
\eea
and $\mathbf{x}$ is the vector with the free parameters to be fit. Note that the factor $c/H_0$ can be reabsorbed into $\mu_0$. Then, one can construct $\Delta\mathcal{F}(\mathbf{x})=\mathcal{F}_{\text{theo}}-\mathcal{F}_{\text{obs}}$, using for this purpose the  distance modulus $\mathcal{F}_{\text{obs}}$ associated with the observed magnitude. At this point it may be thought  that a possible $\chi_{SN}^2$ is
\bea
\chi_{SN}^2(\mathbf{x})&=&\left(\Delta\mathcal{F}(\mathbf{x})\right)^{T}\cdotp C_{SN}^{-1}\cdotp \Delta\mathcal{F}(\mathbf{x}).
\eea

However, the latter would contain the nuisance parameter $\mu_0$, which in turn is a function of the Hubble constant, the speed of light $c$ and the SNeIa absolute magnitude. In order to circumvent this problem, $\chi_{SN}^2$ is marginalized analytically with respect to $\mu_0$ as in \cite{Conley:2011ku}, thus obtaining a new $\chi_{SN}$ estimator of the form:
\bea
\chi_{SN}^2(\mathbf{x})&=&\left(\Delta\mathcal{F}(\mathbf{x})\right)^{T}\cdotp C_{SN}^{-1}\cdotp \Delta\mathcal{F}(\mathbf{x})+\ln{\frac{S}{2\pi}}-\frac{k^2(\mathbf{x})}{S},\nonumber\\
\eea
where $C_{SN}$ is the total covariance matrix, $S$ is the sum of all entries of $C_{SN}^{-1}$, which gives and estimation of the precision of these data are it is independent of $\mathbf{x}$, and $k$ is $\Delta\mathcal{F}(\Omega_m,\Omega_r,\Omega_\Lambda)$ but weighed by a covariance matrix with as follows:
\bea
k(\mathbf{x})={\left(\Delta\mathcal{F}(\mathbf{x})\right)^{T}\cdotp C_{SN}^{-1}}.
\eea

\subsection{Hubble data}

Early time passively evolving galaxies have some peculiar features in their spectra which have been shown to correlate with their evolving stage. Thus,  direct astrophysical measurements can  estimate their differential ages at different redshifts and this can be finally related to the Hubble parameter. For more details, see \cite{Jimenez:2001gg, Moresco:2010wh, Moresco:2012jh}.
Therefore, and essentially, this is  a sample of $31$ values of $H(z)$ for $0.07<z<1.965$ \cite{Moresco:2016mzx, Moresco:2015cya} to assist us  in fitting the free parameters of our theoretical setting through the construction of a $\chi_H^2$ as follows:
\bea
\chi_H^2=\sum_{i=1}^{31}\frac{\left[H\left(z_i,\mathbf{x}\right)-H_{\textit{obs}}(z_i)\right]^2}{\sigma^2_H(z_i)},
\eea
where $H_{\textit{obs}}(z_i)$ is the observed value at $z_i$, $\sigma_H(z_i)$ are the observational errors, and $H\left(z_i,\mathbf{x}\right)$ is the value of a theoretical $H$ for the same $z_i$ with the specific parameter vector $\mathbf{x}$. Needless to say,  for our case of interest $H\left(z_i,\mathbf{x}\right)$ will be given by the positive branch of Eq.~\eqref{Hfinal}.

\subsection{Cosmic Microwave Background data}

It is common practice to condense cosmic microwave background (CMB) data into the so-called shift parameters \cite{Wang:2015tua} when examining  the evolution of the cosmological background. This set of three quantities basically informs us of the position of the first peak in the temperature angular power spectrum through the ratio between its position in the model one wants to analyze and
that of an SCDM model (standard cold dark matter). The set of shift parameters is formed by the exact expression of that quadrature ratio (an approximate yet quite accurate expression) and the normalized density fraction of baryons. We construct the $\chi^2_{CMB}$ estimator  as
\bea
\chi^2_{CMB}=(\Delta\mathcal{F}^{CMB})^{T}\cdotp C_{CMB}^{-1}\cdotp\Delta\mathcal{F}^{CMB},
\eea
where $\Delta\mathcal{F}^{CMB}$ is a vector formed by those three quantities mentioned above \cite{Wang:2015tua}.

We have used the \textit{Planck} 2018 data release \cite{Zhai:2018vmm} to obtain the shift parameters which, according to our earlier sketch,   are defined as:
\bea
&&R(\mathbf{x})\equiv\sqrt{\Omega_m H_0^2}\frac{r(z_*,\mathbf{x})}{c},\\
&&l_a(\mathbf{x})\equiv\pi\frac{r(z_*,\mathbf{x})}{r_s(z_*,\mathbf{x})},\\
&&\omega_b\equiv\Omega_b h^2,
\eea
where $r(z,\mathbf{x})$ is the comoving distance to $z$, which reads
\bea
r(z,\mathbf{x})=\int^{z}_{0}\frac{c}{H(z',\mathbf{x})}dz',
\eea
and $r_s(z_,\mathbf{x})$ is the comoving sound horizon, defined as
\bea
r_s(z_,\mathbf{x})=\int^{\infty}_{z}\frac{c_s(z')}{H(z',\mathbf{x})}dz',
\label{soundhorizon}
    \eea
  In the latter we must take into account that the sound speed $c_s(z)$ is given by the expression
    \bea
    c_s(z)=\frac{c}{\sqrt{3(1+\hat{R}_b(1+z)^{-1})}},
    \eea
    where
    \bea
    \hat{R}_b=31500\Omega_b h^2\left(\frac{T_{CMB}}{2.7}\right)^{-4}.
    \eea
The comoving distance, and the comoving sound horizon are evaluated at the photon-decoupling redshift $z_*$, calculated in appendix E of \cite{Hu:1995en} with
\bea
z_*=1048\left[1+0.00124(\Omega_b h^2)^{-0.738}\right]\left[1+g_1(\Omega_m h^2)^{g_2}\right],\nonumber
\eea
where
\bea
&&g_1=0.0783(\Omega_b h^2)^{-0.238}\left[1+39.5(\Omega_b h^2)^{-0.763}\right]^{-1},\nonumber\\
&&g_2=0.560\left[1+21.1(\Omega_b h^2)^{1.81}\right]^{-1}.\nonumber
\eea

Let us recall that the shift parameters depend on the position of the CMB acoustic peaks, which are functions of the geometry of the  model considered. For that reason, they can be used to discriminate between different models or different values of the free parameters $\mathbf{x}$, which includes $\Omega_b$ in this case. A complete and detailed description of these parameters and those that follow can be found in \cite{Zhai:2018vmm, Komatsu:2008hk}.

\subsection{Baryon Acoustic Oscillations data}

The last set of data addresses Baryon Acoustic Oscillations (BAO), which are fluctuations in the density of visible baryonic matter as a consequence of acoustic density waves in the primordial plasma. Accordingly, there is a distance associated with the maximum distance that acoustic waves can travel through this media until the plasma cooled at the recombination moment, where it became a soup of neutral atoms and the expansion of  plasma density waves
stopped and they got frozen. That being so, the mentioned distance can be used as a standard ruler.

We will use five sets of data, collected by different observational missions. Let us give
now relevant details.
\begin{itemize}
    \item WiggleZ: these are data coming from the WiggleZ Dark Energy survey \cite{Blake:2011en}, which are evaluated at redshifts $z_{\textit{w}}=(0.44,0.66,0.73)$ as shown in Table 1 of \cite{Blake:2012pj} . Following that work, we will consider two quantities: the acoustic parameter given by
    \bea
    A(z,\mathbf{x})=100\sqrt{\Omega_m h^2}\frac{D_V(z,\mathbf{x})}{cz},
    \eea
    and the Alcock-Paczynski distortion parameter:
    \bea
    F(z,\mathbf{x})=(1+z)\frac{D_{A}(z,\mathbf{x})H(z,\mathbf{x})}{c},
    \eea
    where $D_{A}$ is the angular diameter distance
    \bea
    D_{A}(z)=\frac{1}{1+z}\int_0^z\frac{c\;dz'}{H_0 E(z',\mathbf{x})},
    \eea
    and $D_V$ is the geometric mean of the longitudinal ($D_A$) and radial ($c/H(z))$ BAO modes, defined as
    \bea
    D_{V}(z,\mathbf{x})=\left[(1+z)^2D_A^2(z,\mathbf{x})\frac{cz}{H(z,\mathbf{x})}\right]^{1/3}.
    \eea
    Consequently, we have two observational parameters, i.e. $A(z_{\textit{w}})$ and $F(z_{\textit{w}})$ which can be compared with the theoretical value drawn from the model under study with a specific $\mathbf{x}$, and allowing us to construct a new $\Delta\mathcal{F}_{\textit{w}}$. In this cases we define $\chi_{\textit{w}}^2$  as
    \bea
    \chi^2_{w}=(\Delta\mathcal{F}_{\textit{w}})^{T}\cdotp C_{\textit{w}}^{-1}\cdotp\Delta\mathcal{F}_{\textit{w}},
    \eea
    where $C_{\textit{w}}^{-1}$ is a matrix given in Table 2  of \cite{Blake:2012pj}.
    \item BOSS: in this case, we consider the data from the SDSS-III Baryon Oscillation Spectroscopy Survey (BOSS) DR12  described in \cite{Alam:2016hwk}. We proceed analogously to
    the WiggleZ case but now we have $z_{B}=(0.38,0.51,0.61)$, whereas the fundamental parameters are
    \bea
    D_M(z,\mathbf{x})\frac{r_s^{fid}(z_d)}{r_s(z_d,\mathbf{x})},\;\;\;\;\;\;\;\;\;\;\;H(z,\mathbf{x})\frac{r_s(z_d,\mathbf{x})}{r_s^{fid}(z_d)},
    \eea
    where $D_M(z)=r(z)$ is the comoving distance, $r_s(z_d,\mathbf{x})$ denotes the sound horizon defined as Eq.~\eqref{soundhorizon} but evaluated at the dragging redshift $z_d$ and $r_s^{fid}(z_d)$ is the same parameter but calculated for a given fiducial cosmological model which in this specific case is equal to $147.78$ Mpc.
    Clearly, the first step involves  calculating the redshift of the drag epoch $z_d$, which can be done considering the approximation \cite{Eisenstein:1997ik}
    \bea
    z_d=\frac{1925(\Omega_m h^2)^{0.251}}{1+0.659(\Omega_m h^2)^{0.828}}\left[1+b_1(\Omega_b h^2)^{b_2}\right],
    \eea
    where $b_1$ and $b_2$ are factors calculated as follows:
    \bea
    b_1&=&0.313(\Omega_m h^2)^{-0.419}\nonumber\\
    &&\times\left[1+0.607(\Omega_m h^2)^{0.6748}\right],\\
    b_2&=&0.238(\Omega_m h^2)^{0.223}.
    \eea
    Once more, we will define:
     \bea
    \chi^2_{B}=(\Delta\mathcal{F}_{\textit{B}})^{T}\cdotp C_{\textit{B}}^{-1}\cdotp\Delta\mathcal{F}_{\textit{B}},
    \eea
    where $\Delta\mathcal{F}_{\textit{B}}$ is the difference between the observational data and the resulting value for $\mathbf{x}$, and $C_{\textit{B}}^{-1}$ is the inverse of the covariance matrix given in Table 8 of \cite{Alam:2016hwk}.
    \item eBOSS: the extended Baryon Oscillation Spectroscopy Survey (eBOSS) gives us one more data point \cite{Ata:2017dya}: \bea D_V(z=1.52)=3843\pm147\frac{r_s(z_d)}{r_s^{fid}(z_d)}\, {\rm Mpc}. \eea 
    Our function $\chi^2_{eB}$ gets a simpler expression that in other cases, as the matrix notation is not necessary, and we simply have
    \bea
    \chi^2_{eB}=\frac{\Delta\mathcal{F}^2_{eB}}{\sigma_{eB}^2}
    \eea
    where $\sigma_{eB}^2$ is the error in the datum.
    \item BOSS-Lyman $\alpha$: another set of data is Quasar-Lyman $\alpha$ Forest from SDSS-III BOSS DR11 \cite{Agathe:2019vsu} which contributes two new data points to the analysis:
    \bea
    \frac{D_M(z=2.34,\mathbf{x})}{r_s(z_d,\mathbf{x})}=36.98^{+1.26}_{-1.18},\\
    \frac{c}{H(z=2.34,\mathbf{x})r_s(z_d,\mathbf{x})}=9.00^{+0.22}_{-0.22},
    \eea
    and its $\chi^2$ is defined as usual.
    \item Finally, we consider the voids-galaxy cross-correlation data from \cite{Nadathur:2019mct}. This set gives us two new data points at $z=0.57$ which are
    \bea
    &&\frac{D_A(z=0.57,\mathbf{x})}{r_s(z_d,\mathbf{x})}=9.383\pm 0.077,\\
    &&H(z=0.57,\mathbf{x})r_s(z_d,\mathbf{x})=(14.05\pm 0.14) \frac{10^3\,\mathrm{km}}{\mathrm{s}}
    \eea
    and the same usual definition applies to its $\chi^2$.
\end{itemize}

\section{Results}\label{Results}

{\renewcommand{\tabcolsep}{1.5mm}
{\renewcommand{\arraystretch}{1.5}
\begin{table*}[htbp]
\caption{MCMC best fits and errors. Quantities in italic correspond to secondary  parameters.}
\begin{tabular}{ccccccc}
\hline
\hline
& & Pantheon & Hubble & CMB & BAO & Total  \\
\hline \\
\multirow{3}{*}{$\Omega_m$} & $\Lambda$CDM &  $0.298_{-0.021}^{+0.022}$   & $0.327_{-0.056}^{+0.066}$ & $0.316_{-0.007}^{+0.007}$& $0.320_{-0.015}^{+0.016}$& $0.323_{-0.005}^{+0.005}$\\
\multicolumn{1}{l}{}       &    $f(Q)_{\Omega_\Lambda\neq0}$   &   $0.337_{-0.073}^{+0.075}$   &  $0.341_{-0.060}^{+0.070}$ &   $0.346_{-0.080}^{+0.092}$   & $0.323_{-0.017}^{+0.020}$ &   $0.325_{-0.007}^{+0.007}$   \\
\multicolumn{1}{l}{}       &    $f(Q)_{\Omega_\Lambda=0}$   &  $0.400_{-0.024}^{+0.024}$  &  $ 0.350_{-0.049}^{+0.057}$ & $0.238_{-0.006}^{+0.006}$ &
$0.348_{-0.016}^{+0.016}$ & $0.285_{-0.004}^{+0.004}$  \\
\hline
\multicolumn{1}{c}{\multirow{3}{*}{$\Omega_b$}} & $\Lambda$CDM & - & -  &$0.0491_{-0.0006}^{+0.0006}$ & $0.063_{-0.031}^{+0.012}$ & $0.0496_{-0.0004}^{+0.0004}$\\
\multicolumn{1}{l}{}  & $f(Q)_{\Omega_\Lambda\neq0}$ & -  & - &  $0.057_{-0.012}^{+0.014}$ &  $0.081_{-0.036}^{+0.019}$&    $0.0501_{-0.0010}^{+0.0010}$       \\
\multicolumn{1}{l}{}       &    $f(Q)_{\Omega_\Lambda=0}$   &   -   &  - &
$0.0371_{-0.0005}^{+0.0005}$  & $0.042_{-0.021}^{+0.011}$ &  $0.0407_{-0.0004}^{+0.0004}$ \\
\hline
\multicolumn{1}{c}{\multirow{3}{*}{$h$}} & $\Lambda$CDM &  $-$  & $0.678_{-0.031}^{+0.031}$ & $0.675_{-0.005}^{+0.005}$& $>0.65$ & $0.670_{-0.003}^{+0.003}$ \\  \multicolumn{1}{l}{}       &    $f(Q)_{\Omega_\Lambda\neq0}$     & $-$  &   $0.674_{-0.054}^{+0.039}$    &   $0.645_{-0.071}^{+0.090}$  & $>0.62$ &   $0.667_{-0.007}^{+0.007}$     \\
\multicolumn{1}{l}{}       &    $f(Q)_{\Omega_\Lambda=0}$   & $-$   &  $0.703_{-0.030}^{+0.029}$ &  $0.777_{-0.007}^{+0.007}$  & $>0.70$ &  $0.730_{-0.004}^{+0.004}$ \\ 
\hline
\multicolumn{1}{c}{\multirow{3}{*}{$\Omega_\Lambda$}} & $\Lambda$CDM &
$\mathit{0.701_{-0.022}^{+0.021}}$ &  $\mathit{0.673_{-0.066}^{+0.056}}$ & $\mathit{0.684_{-0.007}^{+0.007}}$ & $\mathit{0.680_{-0.016}^{+0.015}}$ & $\mathit{0.677_{-0.005}^{+0.005}}$ \\
\multicolumn{1}{l}{}    &    $f(Q)_{\Omega_\Lambda\neq0}$        & $0.43_{-0.49}^{+0.47}$  & $0.64_{-0.60}^{+0.59}$ & $0.87_{-0.57}^{+0.43}$ & $1.11_{-0.18}^{+0.21}$ &       $0.701_{-0.053}^{+0.054}$       \\
\multicolumn{1}{l}{}       &    $f(Q)_{\Omega_\Lambda=0}$   &   -   &  $-$ &  $-$  & $-$ &  $-$  \\
\hline
\multicolumn{1}{c}{\multirow{3}{*}{$\Omega_Q$}} & $\Lambda$CDM & -  & $-$  & $-$ & $-$ & $-$ \\
\multicolumn{1}{l}{}  & $f(Q)_{\Omega_\Lambda\neq0}$  & $\mathit{0.23_{-0.40}^{+0.42}}$  & $\mathit{0.03_{-0.61}^{+0.58}}$ & $\mathit{-0.22_{-0.52}^{+0.65}}$  & $\mathit{-0.43_{-0.22}^{+0.18}}$ &       $\mathit{-0.027_{-0.058}^{+0.057}}$    \\
\multicolumn{1}{l}{}       &    $f(Q)_{\Omega_\Lambda=0}$   &   $\mathit{0.599_{-0.024}^{+0.023}}$   &  $\mathit{0.650_{-0.057}^{+0.049}}$ &  $\mathit{0.762_{-0.006}^{+0.006}}$  & $\mathit{0.651_{-0.016}^{+0.016}}$ &  $\mathit{0.715_{-0.004}^{+0.004}}$ \\
\hline 
\hline
\multicolumn{1}{c}{\multirow{3}{*}{$\chi^2$}} & $\Lambda$CDM & 1035.77 & 14.49  &  0.001 &  16.55  &  1072.19 \\
\multicolumn{1}{l}{}                          & f(Q)$_{\Omega_\Lambda\neq0}$   & 1035.72 & 14.40 & 0.005 & 11.34 & 1072.01  \\ 
\multicolumn{1}{l}{}                          &    f(Q)$_{\Omega_\Lambda=0}$   & 1036.48 & 14.53& 0.003 & 51.34& 1207.96 \\
\hline
\multicolumn{1}{c}{\multirow{3}{*}{$\mathcal{B}^{i}_{j}$}} & $\Lambda$CDM & - & - & - & - & 1 \\
\multicolumn{1}{l}{}                                      & f(Q)$_{\Omega_\Lambda\neq0}$ & - & - & - & - & 0.76\\
\multicolumn{1}{l}{}                                      & f(Q)$_{\Omega_\Lambda=0}$ & - & - & - & - & $3\cdot10^{-30}$ \\
\hline
\multicolumn{1}{c}{\multirow{3}{*}{$\ln \mathcal{B}^{i}_{j}$}} & $\Lambda$CDM & - & - & - & - & 0 \\
\multicolumn{1}{l}{}                                      & f(Q)$_{\Omega_\Lambda\neq0}$ & - & - & - & - & $-0.27$\\
\multicolumn{1}{l}{}                                      & f(Q)$_{\Omega_\Lambda=0}$ & - & - & - & - & $-68$ \\
\hline
\end{tabular}
\label{tableresults}
\end{table*}}}

After having  set forth our statistical procedure along with a detailed outline of the astrophysical probes chosen, we can now present the results of our research.

We stress once more that a key aspect  is the comparison of the novel $f(Q)$ scenario  under consideration with the $\Lambda$CDM standard. To this end we  simply perform all the tests on this scenario as well.

From a very wide perspective, we could also wonder about the peculiar case of our  $f(Q)$ model that follows from setting $\Omega_\Lambda=0$. In principle, this  scenario  would be of interest as to  avoid  the problems associated with the cosmological constant. Although some insight into this particular setting can be easily drawn from  the more general $f(Q)$ one, we treat it in full so that our conclusions are more complete and precise. 

The  best fits of (the three) models we analize are shown in Table \ref{tableresults}. Let us emphasize that $\Omega_Q$ is a characteristic parameter of the $f(Q)$ scenario,  a signature of it  that does not appear at all in the $\Lambda$CDM setting.

\begin{figure*}[htb]
    \includegraphics[width=0.4\linewidth]{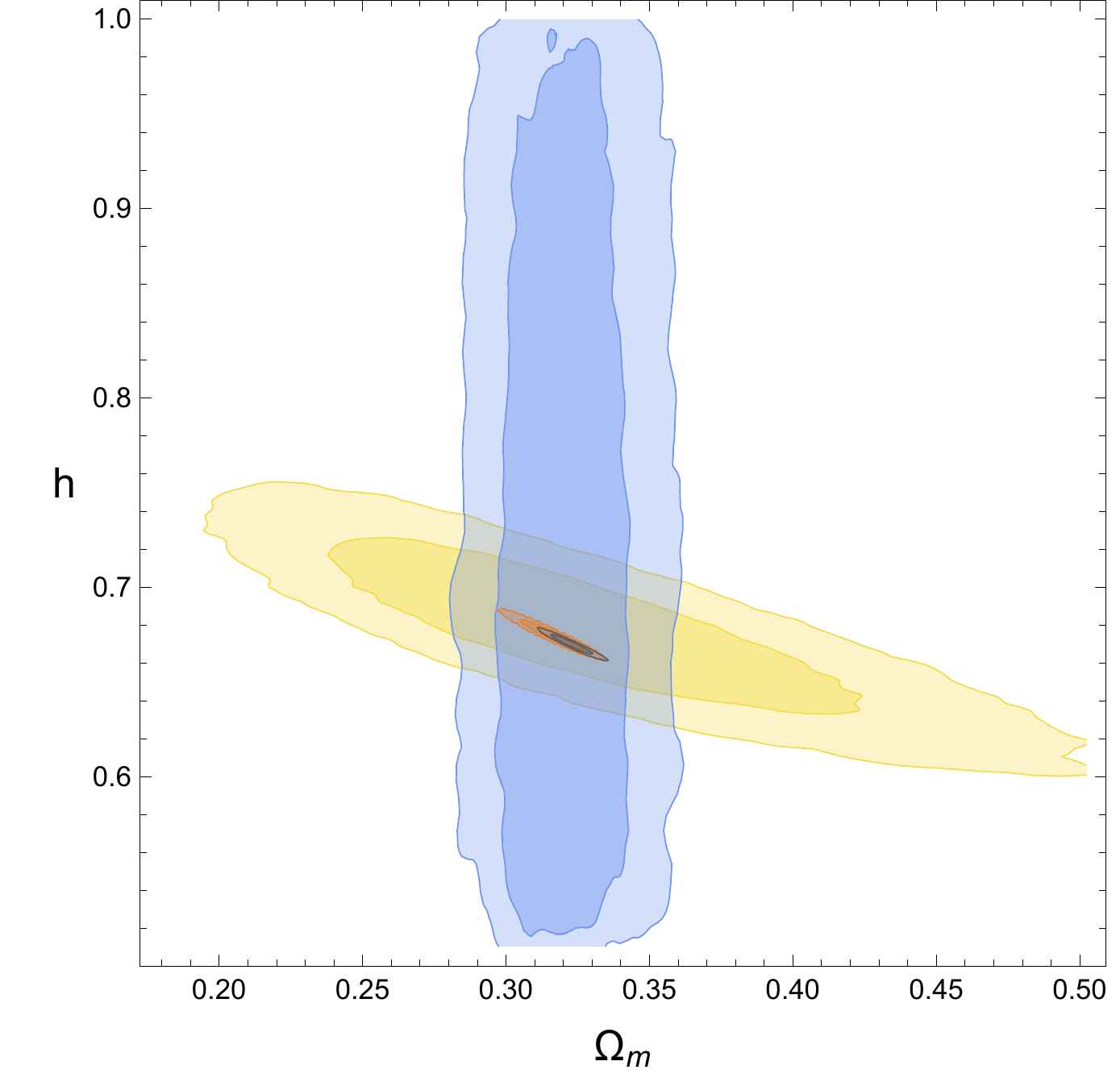}~~~
    \includegraphics[width=0.4\linewidth]{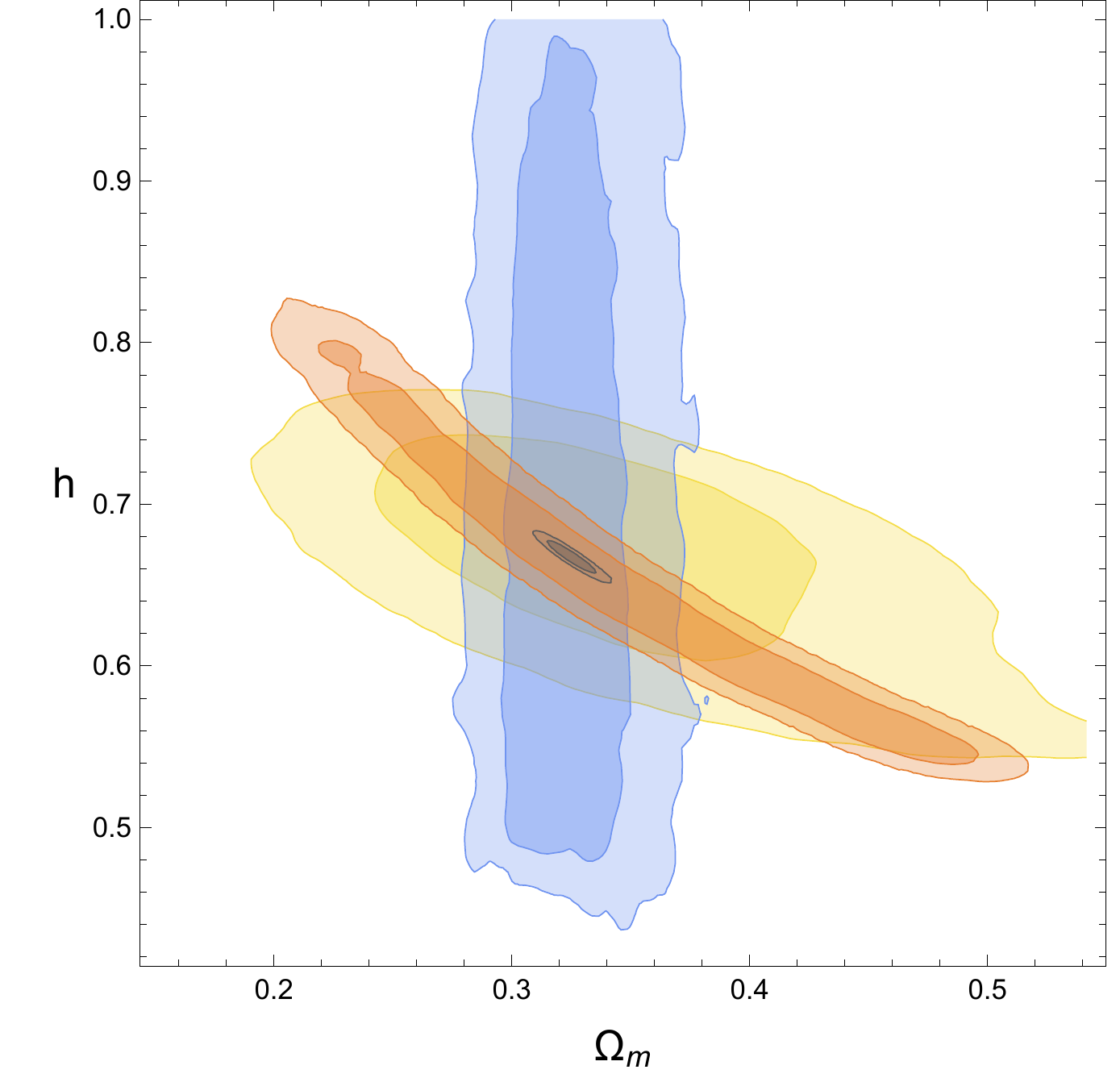}\\
    ~~~\\
    \includegraphics[width=0.4\linewidth]{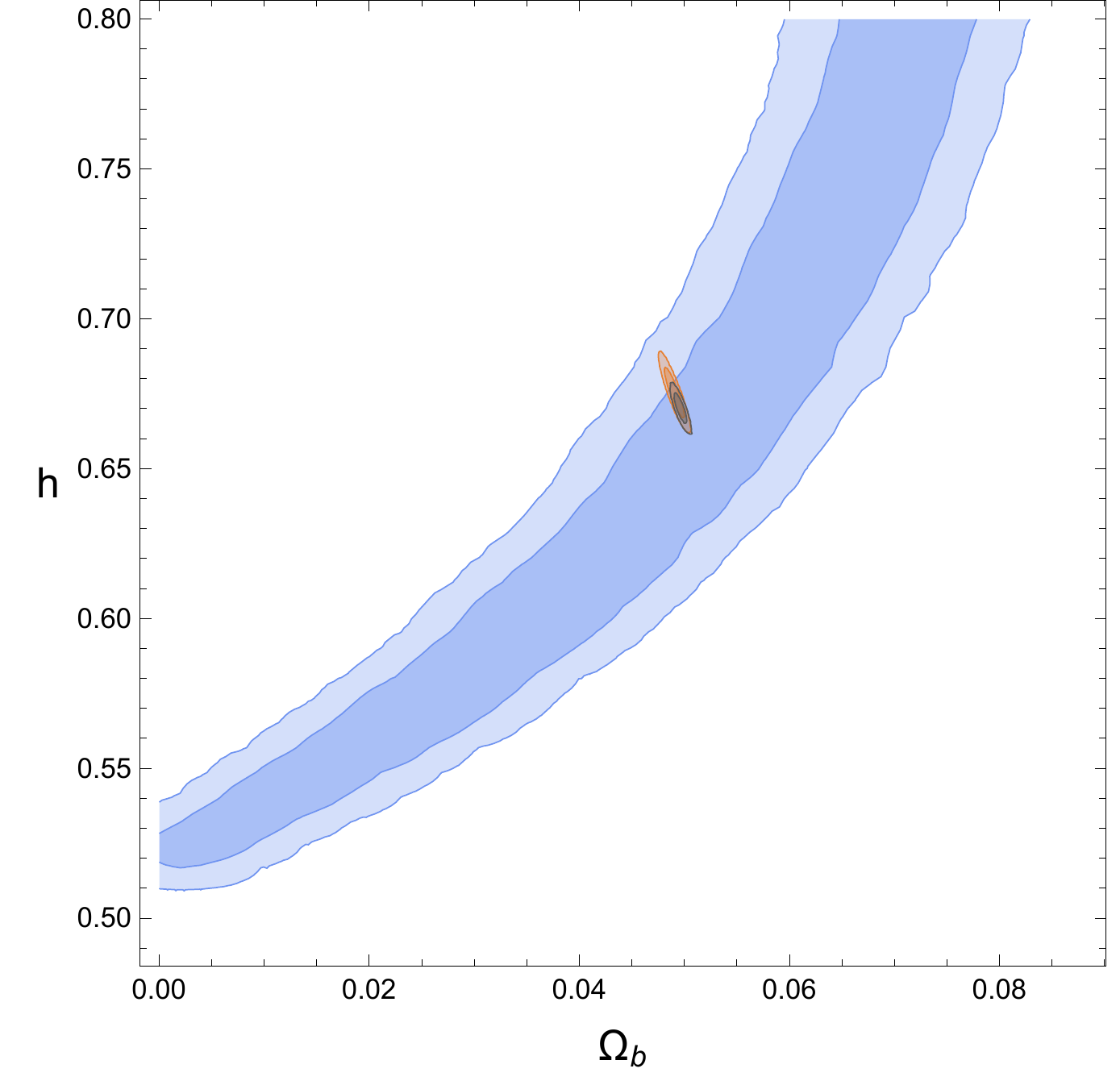}~~~
    \includegraphics[width=0.4\linewidth]{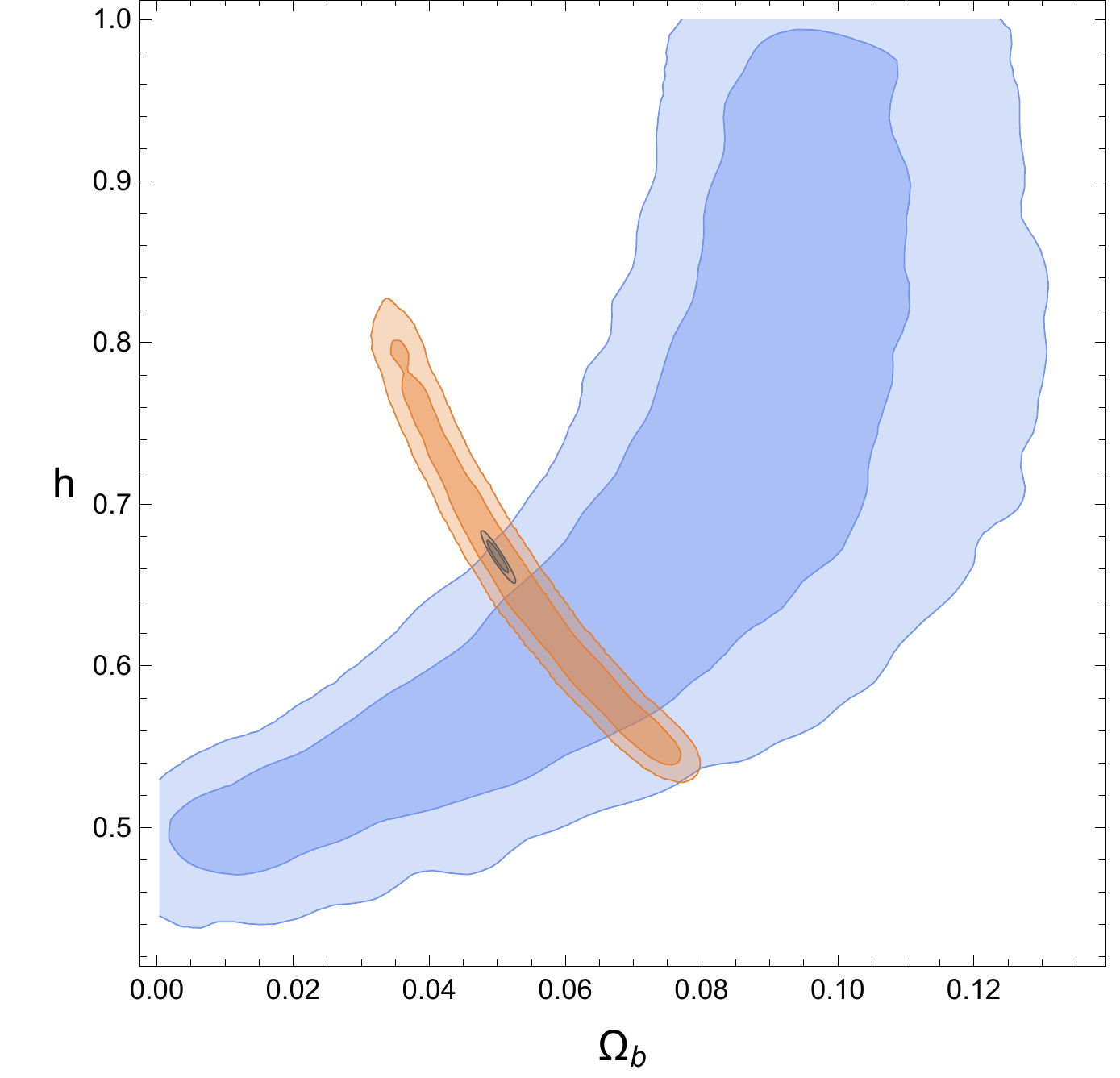}\\
    ~~~\\
    \includegraphics[width=0.4\linewidth]{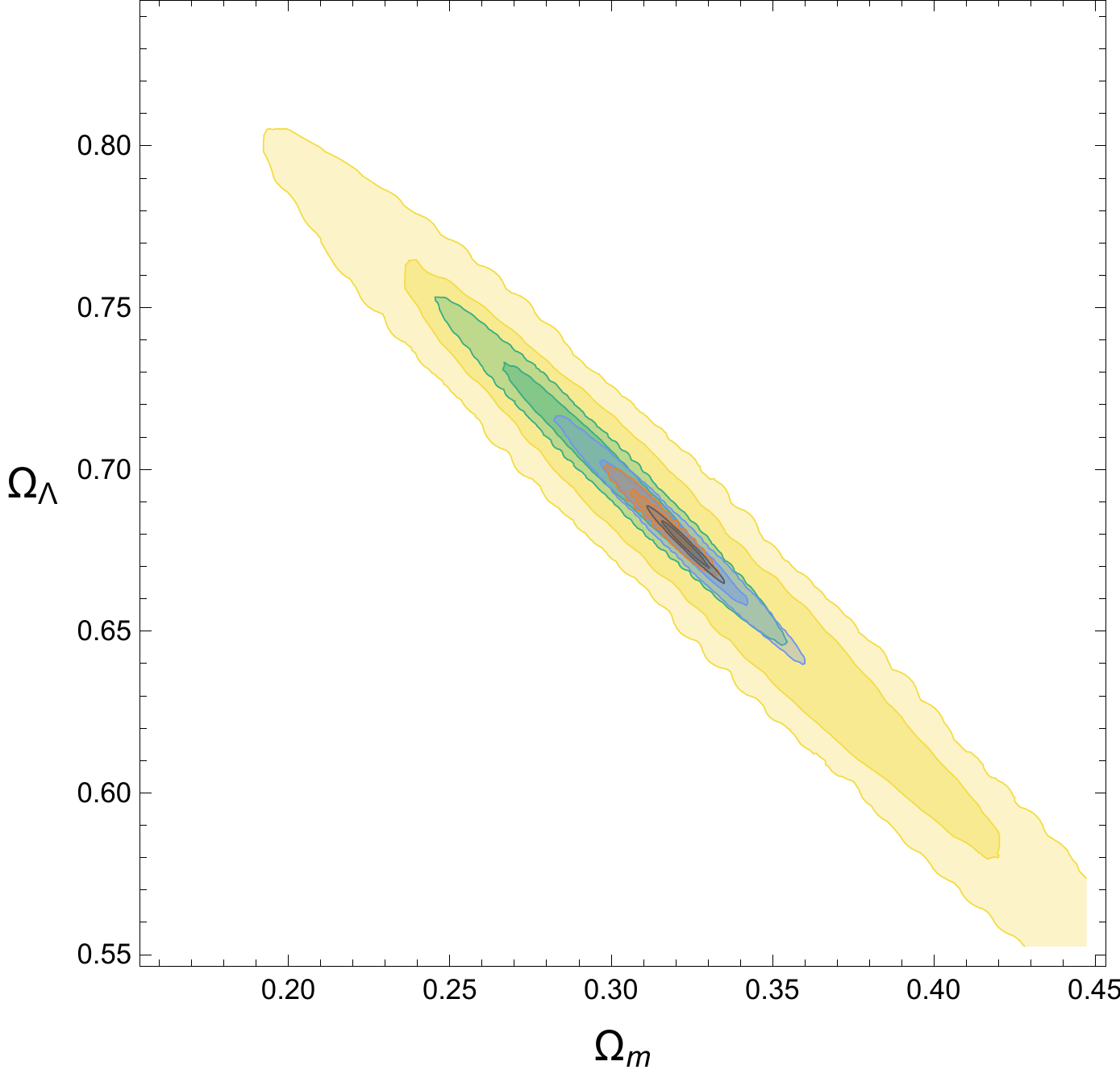}~~~
    \includegraphics[width=0.4\linewidth]{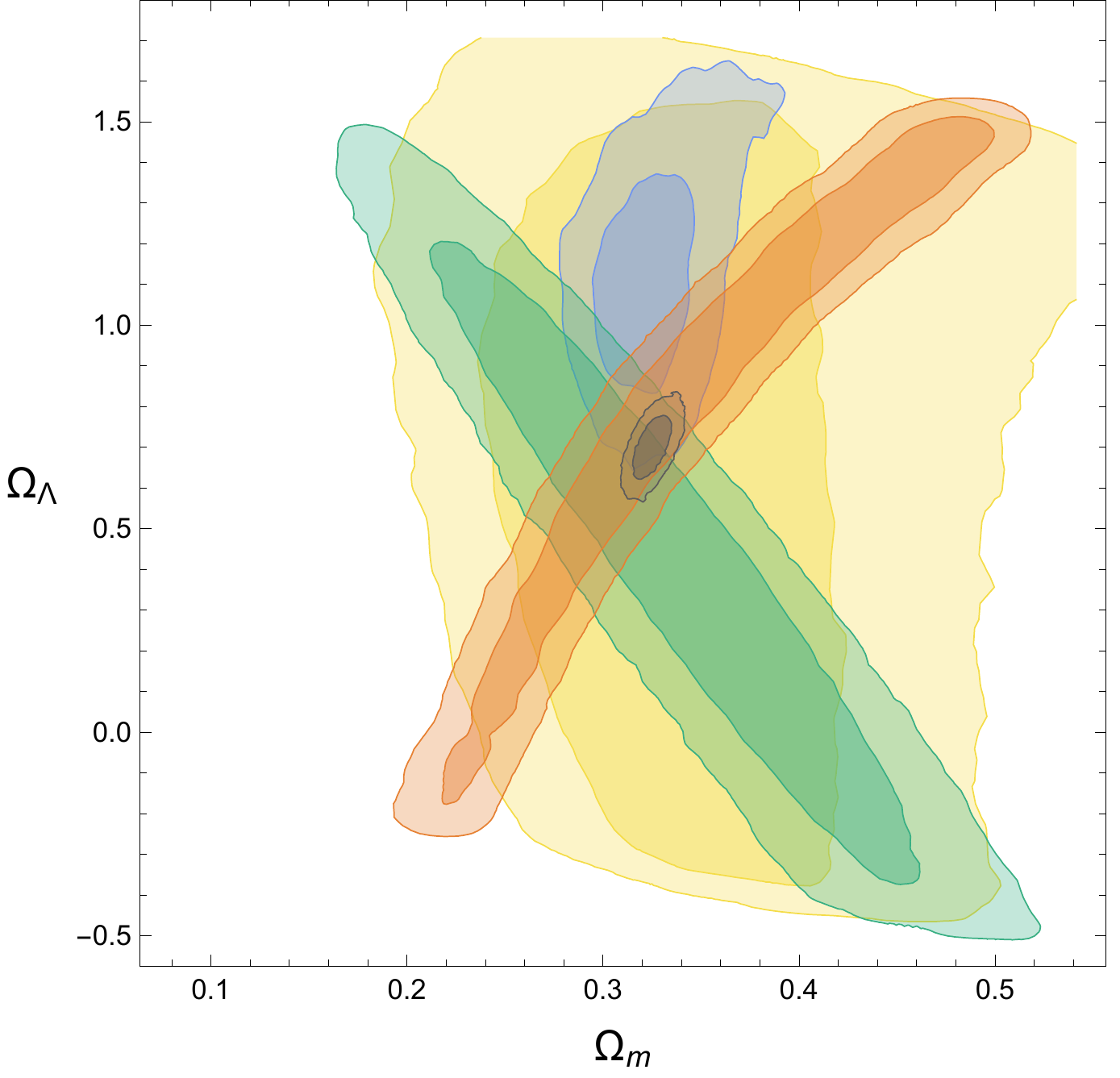}
    \caption{Contour plots for the $\Lambda$CDM model (left column) and the $f(Q)$ model with $\Omega_{\Lambda}\neq 0$ (right column) with the following color scheme: green  - SNeIa, yellow - Hubble data, orange - \textit{Planck 2018} CMB, blue - BAO data, black - all sets of data. As SNeIa are (of course)  unable to  fix the  value $h$ their contours are missing from those plots where constraints on $h$ is shown; for the same rationality, both SNeIa and Hubble contours are absent from plots showing constraints on $\Omega_b$.}
    \label{ContourACDM}
\end{figure*}

\begin{figure*}[htbp]
    \includegraphics[width=0.4\linewidth]{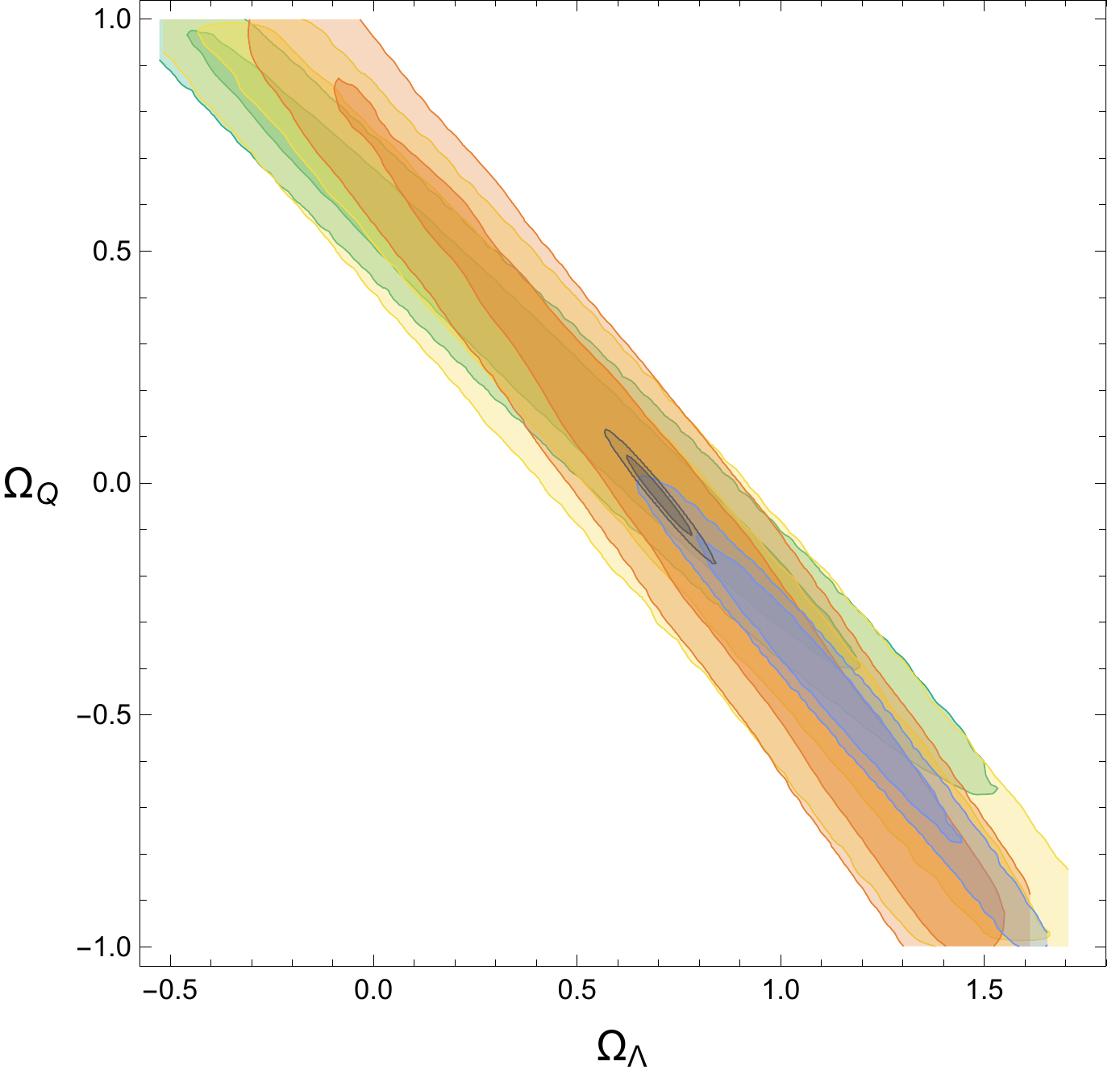}~~~
    \includegraphics[width=0.4\linewidth]{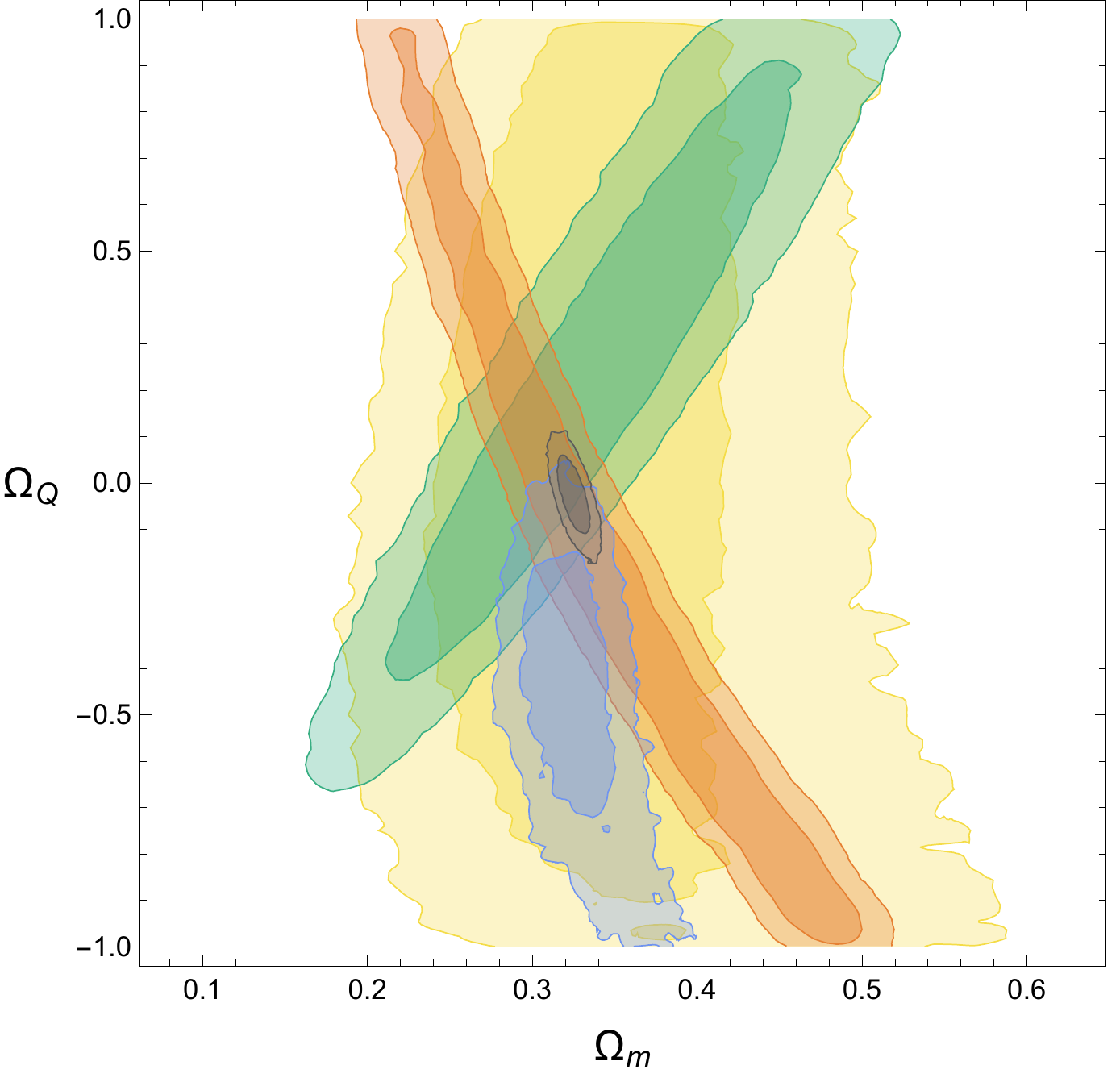}
    \caption{Contour plots for the $f(Q)$ model with $\Omega_{\Lambda} \neq 0$ with the following color scheme: green  - SNeIa, yellow - Hubble data, orange - \textit{Planck 2018} CMB, blue - BAO data, black - all sets of data.}
    \label{ContourfQ}
\end{figure*}

\begin{figure*}[htpb]
\includegraphics[width=0.4\linewidth]{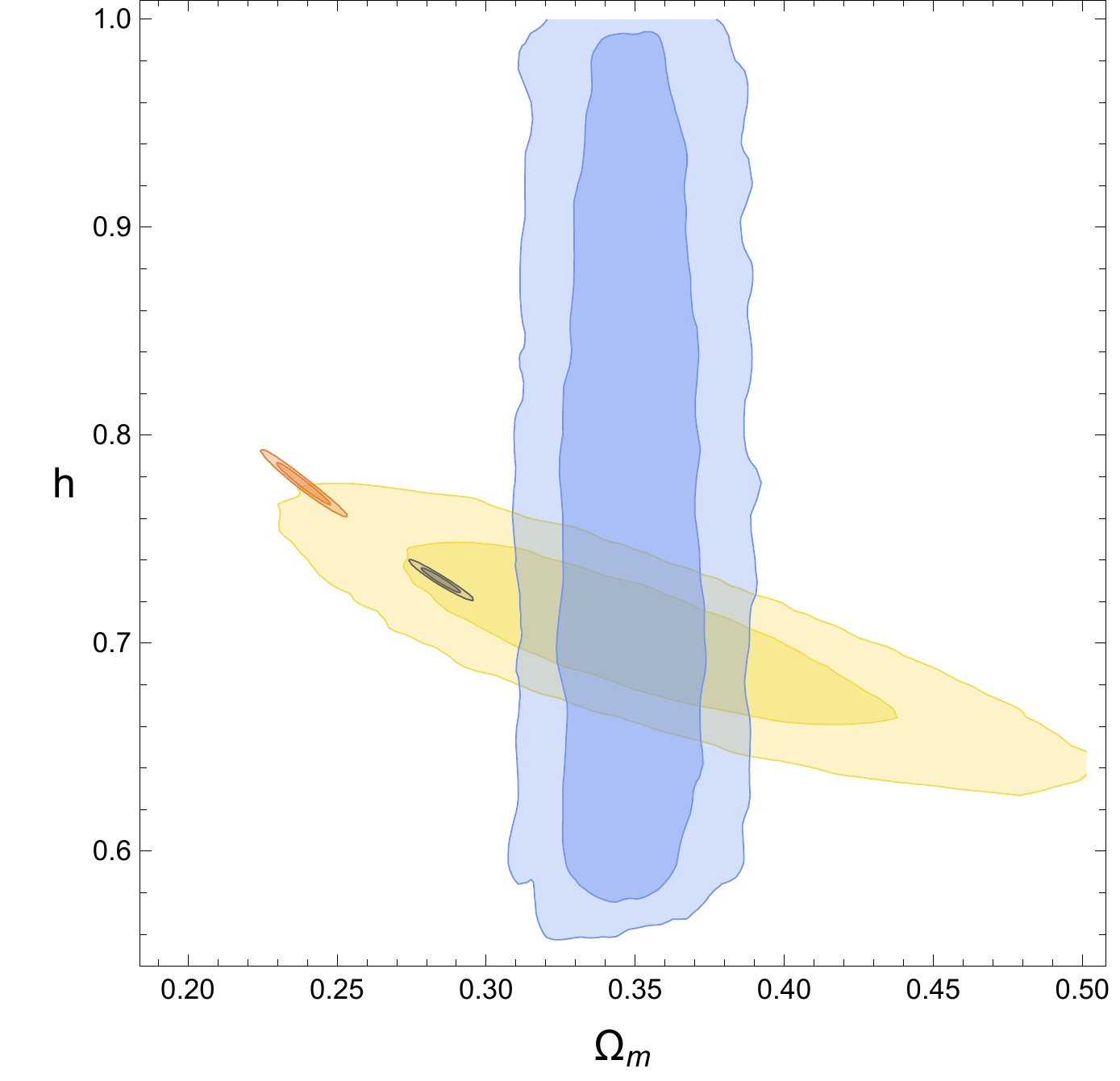}~~~
\includegraphics[width=0.4\linewidth]{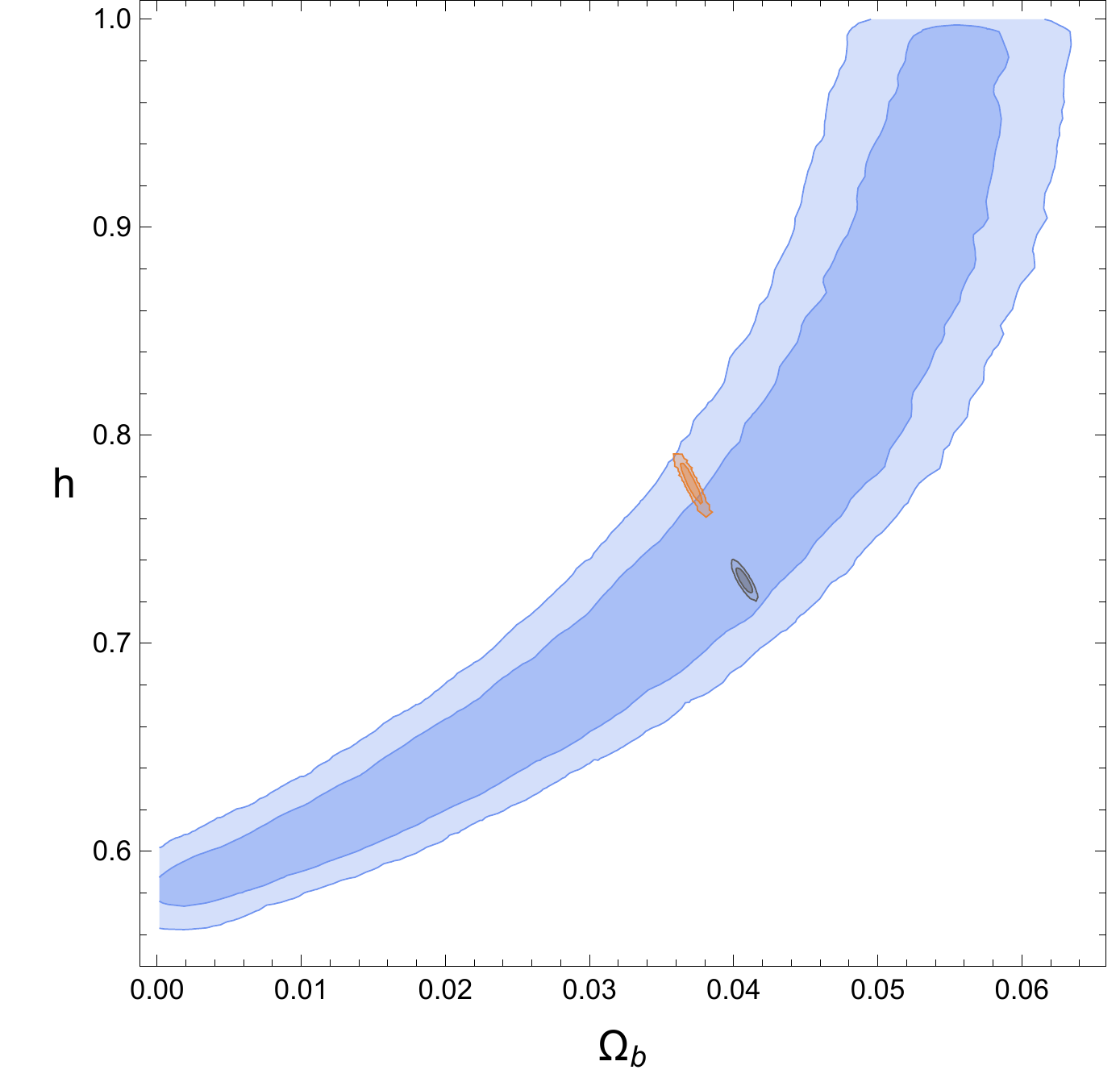}\\
~~~\\
\includegraphics[width=0.4\linewidth]{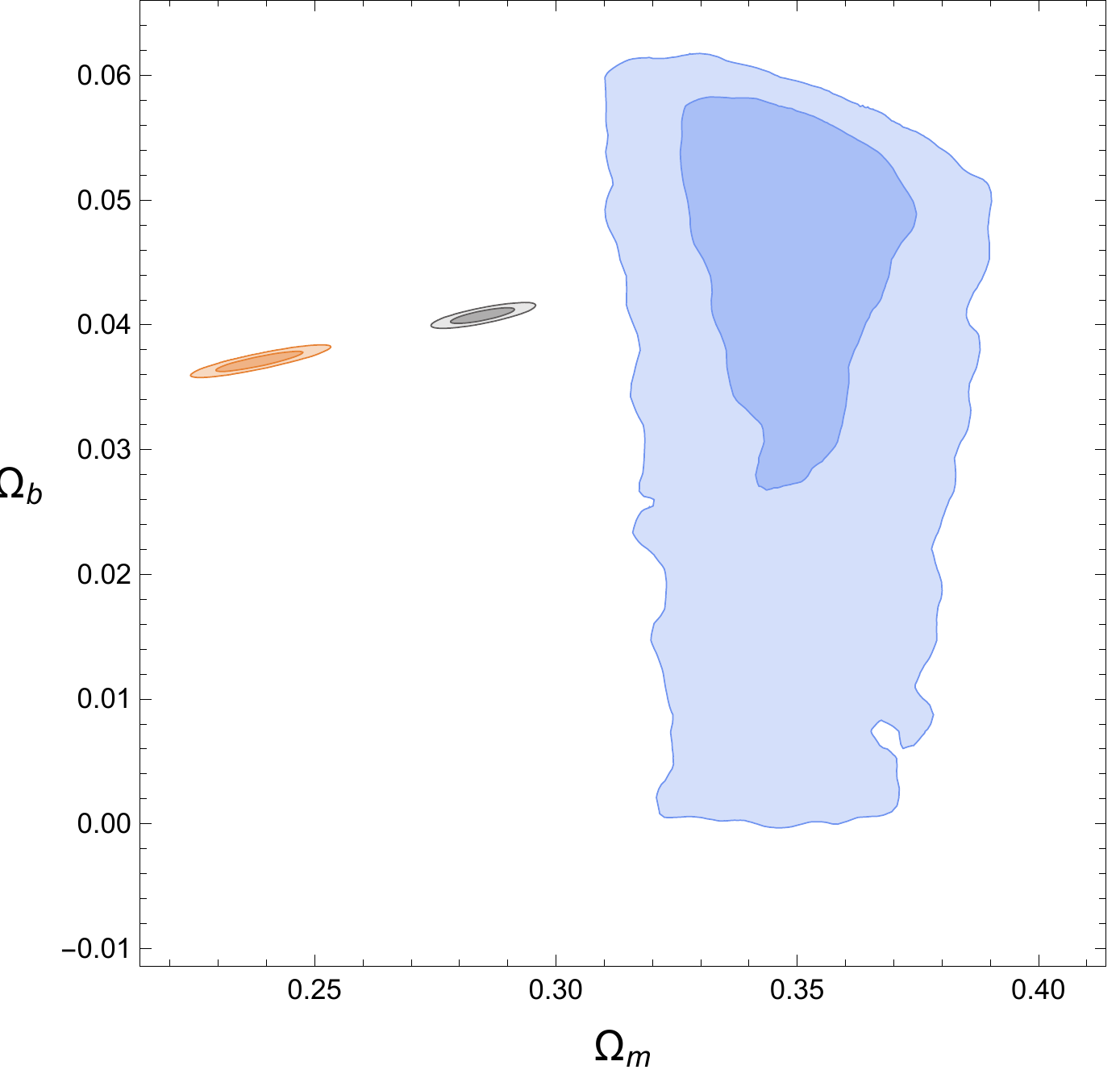}~~~
\includegraphics[width=0.4\linewidth]{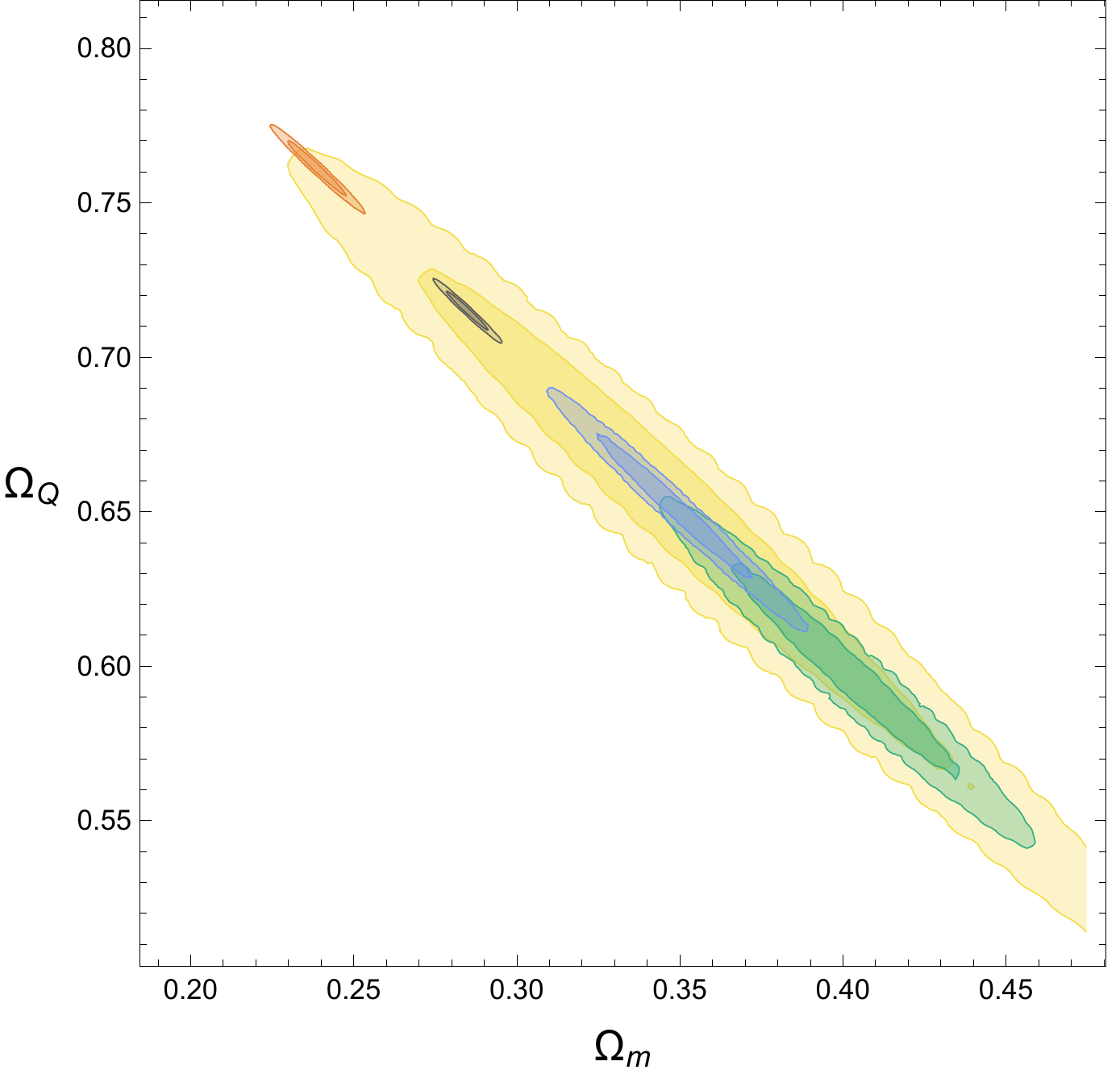}
    \caption{Contour plots for the $f(Q)$ model with $\Omega_{\Lambda} = 0$ using  the following color scheme: green  - SNeIa; yellow - Hubble data; orange - \textit{Planck 2018} CMB; blue - BAO data; black - all sets of data.}
    \label{ContournoA}
\end{figure*}

For additional discernment, we present the marginalized confidence contours directly as drawn from the MCMC procedure providing our best fits. This supplies visual hints of the complementarity of different data sets, their constraining power, and correlation among parameters. The comparison of the contour plots for $\Lambda$CDM and $f(Q)$ with $\Omega_{\Lambda}\neq 0$ can be found in Fig.~\ref{ContourACDM}, while specific results of the $f(Q)$ model are presented separately in Fig.~\ref{ContourfQ}. Finally, results for $f(Q)$ with $\Omega_{\Lambda}=0$ are shown in Fig.~\ref{ContournoA}. For each individual data set or data set combination we draw the contours by choosing two shades of a single colour, and we let the dark and light hues represent the $1\sigma$ and $2\sigma$ regions respectively.

One of the main conclusions of the MCMC analysis is that  in the case of $f(Q)$ with $\Omega_{\Lambda} \neq 0$  the combination of data sets yields a tiny negative value of $\Omega_Q$. Specifically it lies in a significantly wide uncertainty range that makes it compatible with a null value at the $1\sigma$ level (see Table~\ref{tableresults} and the black contour in Fig.~\ref{ContourfQ}). This  compatibility with a vanishing $\Omega_Q$ applies for all the separate data sets except for BAO, which not only bet on a larger negative value only compatible with $\Omega_Q=0$ at the $2\sigma$ level, but also exhibit much lower errors (at least twice smaller) than other probes. The large  uncertainty in $\Omega_Q$ can undoubtedly be associated to that in $\Omega_\Lambda$. It is visually manifest that the two parameters are mutually quite (anti)correlated for all data sets, as the inclination of the contours is very close to $-45^{\circ}$. This significant negative correlation, expected from Eq.~\eqref{omegaqdef}, makes the large degree of uncertainty in the $\Omega_\Lambda$ induce the same behaviour on $\Omega_Q$.

A second outcome is that, independently of the values of $\Omega_\Lambda$ and $\Omega_Q$, the fits of the other two parameters,  $\Omega_m$ and $h$, are very similar in the $f(Q)$ and $\Lambda$CDM scenarios. This  makes sense considering previous arguments, because shifts in $\Omega_\Lambda$ are reabsorbed into $\Omega_Q$ and vice-versa, affecting quite little the rest of parameters. A reflection of this fact are the  very similar  $\chi^2$ values displayed by both models in Table~\ref{tableresults}. 

But we may also note that, although the $\Lambda$CDM and $f(Q)$ with $\Omega_{\Lambda}\neq 0$ models yield similar best fits of $\Omega_m$ and $\Omega_\Lambda$, the distinctive feature encoded in $\Omega_Q$ changes quite significantly the  correlations between those two parameters (see the bottom row of Fig.~\ref{ContourACDM}).

However the behaviour of $\Omega_m$ and $h$ is approximately similar in both models (see top row of Fig.~\ref{ContourACDM}) in a broad sense and in particular in what concerns the correlation among the two parameters. Still there is a noticeable difference, which is the quite larger uncertainty on $\Omega_m$ as associated with CMB data in the case of $f(Q)$ with $\Omega_{\Lambda}\neq 0$. We infer accordingly that the roles of $\Omega_m$ and $h$ seem to be quite similar at low redhifts, but the same does not apply  at high redhifts for what $\Omega_m$ is concerned.

Recalling the Bayes factors, the conclusion is (again) that, within the current data sets, we are not able to distinguish one model from the other. This reasoning is obviously drawn from a background examination, a perturbative one might propound more refined pieces of information (some results along this route were sketched in  \cite{Jimenez:2019ovq}). 

We now may come back the possibility of imposing $\Omega_\Lambda=0$.  Fig. \ref{ContourfQ} provides insight on this matter, as it shows that $\Omega_\Lambda$ is enormously correlated with $\Omega_Q$; for that reason  that parameter could in principle take the role of the cosmological constant and thus give the same evolution, but then the model would then be irreducible to $\Lambda$CDM, as it can be seen by looking at Eq.~\eqref{Hfinal}. However, a look at  the locus on the contours of Fig. \ref{ContourfQ} which corresponds to $\Omega_\Lambda=0$ suggest that this value is clearly disfavoured. Nevertheless, we have decided to perform a direct MCMC analysis to confirm such concerns (see Table~\ref{Cosmographicvalues}).

The  strong tension among all single data sets in this restricted scenario becomes manifest in Fig.~\ref{ContournoA}. 
Paradoxically, the CMB constraints are much better than those derived in the $f(Q)$ case with $\Omega_{\Lambda} \neq 0$, but they require an abnormally low value for $\Omega_m$ which is not consistent with any of the other probes considered. Moreover, although the single $\chi^2$ are comparable with those from other frameworks, we must note that BAO are produce very poor constraints, and that the best fit $\chi^2$ coming from the joint use of all the probes is much larger. If we finally look at the values of the Bayes factors, we can see how this scenario is frankly statistically disfavoured with respect to the other cases.

In addition to these results, we can compute  the cosmographic parameters deceleration $q_0$, jerk $j_0$, and snap $s_0$ to add more elements to the comparison between the kinematics of the two models.  Table~\ref{Cosmographicvalues} summarizes our findings: best fits all share an uncertainty which is much larger in the $f(Q)$ case than in the $\Lambda$CDM case, and with the latter estimations which fall completely within their respective counterparts in the former ones.

\begin{figure*}[ht]
\includegraphics[width=0.4\linewidth]{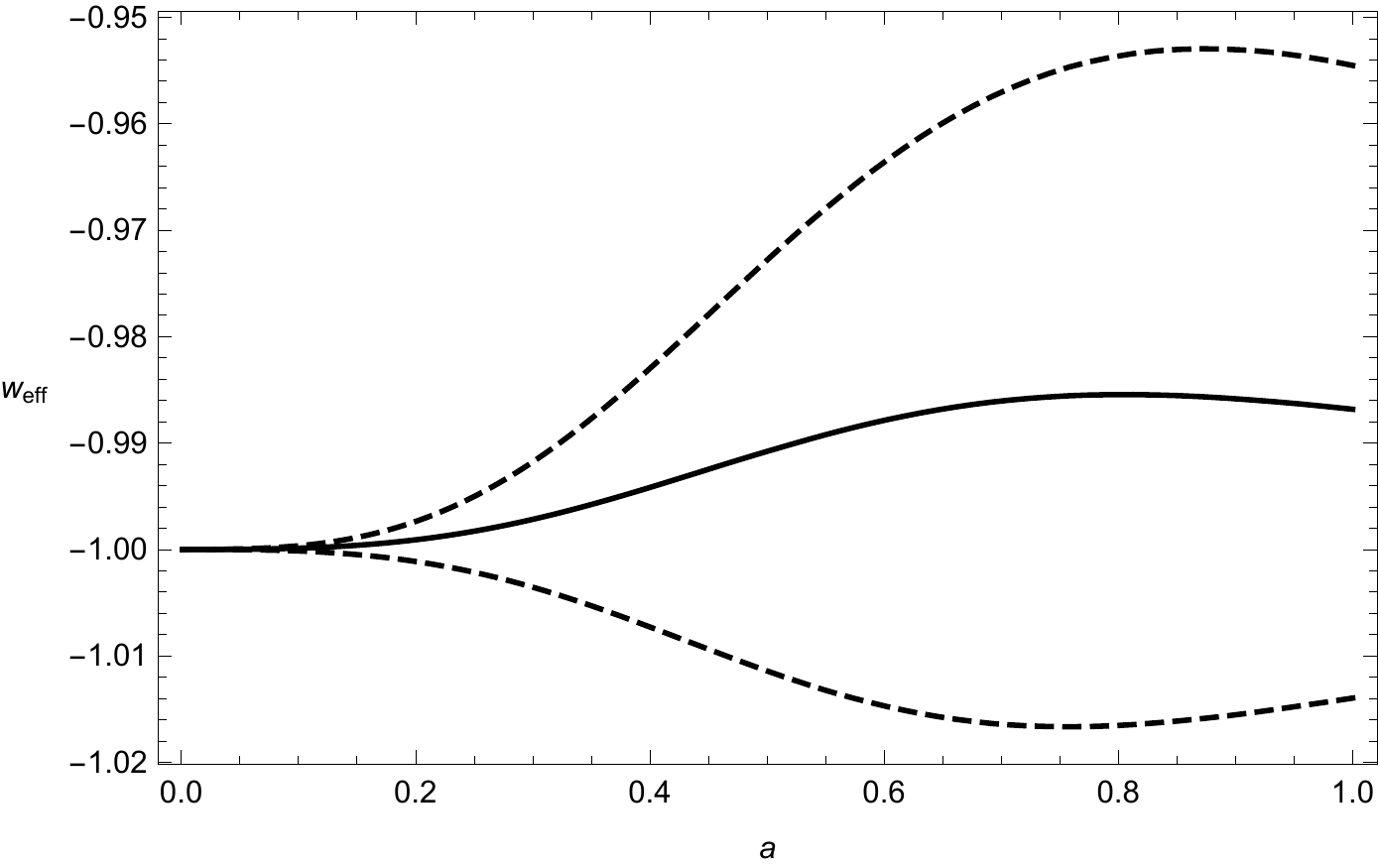}
    \caption{Evolution of $w_{\rm eff}$ as function of $a$. The solid line is the value as drawn from the best fit, whilst the dashed lines mark the boundaries of the confidence interval.}
    \label{weff}
\end{figure*}

{\renewcommand{\tabcolsep}{1.mm}
{\renewcommand{\arraystretch}{1.5}
\begin{table}[htbp]
\caption{Best fits of the cosmographic parameters.}
\begin{tabular}{c ccc}
\hline\hline
    & \multicolumn{1}{c}{$q_0$} &\multicolumn{1}{c}{$j_0$} &\multicolumn{1}{c}{$s_0$}\\ \hline
  $\Lambda$CDM &
 $-0.515_{-0.007}^{+0.007}$ &
$1.000186_{-2\cdot10^{-6}}^{+2\cdot10^{-6}}$ &
  $-0.454_{-0.021}^{+0.021}$
   \\ \cline{1-4}
\multicolumn{1}{l}{}           f(Q)$_{\Omega_\Lambda\neq0}$   &
$-0.499_{-0.035}^{+0.040}$ &
$0.973_{-0.081}^{+0.053}$ &
$-0.453_{-0.039}^{+0.029}$
\\ \hline
\end{tabular}
\label{Cosmographicvalues}
\end{table}}}

Finally, and to close this section, we confront once again our modified gravity scenario with $\Lambda$CDM by rewriting it as a model fuelled by dark matter, radiation and dark energy following for instance \cite{Starobinsky:1998fr,Nakamura:1998mt,Huterer:1998qv,Lazkoz:2007zk}, by setting  
\begin{equation}
p_{\rm eff}=w_{\rm eff}\rho\, ,
\end{equation} and
\begin{equation}
H^2=H_0^2\left[\Omega_m(1+z)^3+H_0^2\Omega_r(1+z)^4\right]+\frac{8\pi G}{3}\rho_{\rm eff}.
\end{equation}
If we combine the last two expressions with the continuity equation, Eq.~\eqref{continuityequation}, we finally write
\bea
w_{\rm eff}(z)=\frac{\displaystyle 2 (1+z)\frac{d\ln E(z)}{dz}-E^{-2}(z)\Omega_r(1+z)^4-3}{3(1-E^{-2}(z)\left[\Omega_m(1+z)^3+\Omega_r(1+z)^4\right])}.\qquad
\eea
The explicit expression of $w_{\rm eff}$ for our modified gravity model is too complicated for it to convey readily usable information, so we regard it sufficient to plot it as a function of  $a$ in Fig. \ref{weff} using the best fit values coming from the MCMC. 
 In addition, and once more with the results of the MCMC analysis, we are able to calculate the value of $w_{\rm eff}$ at the present day for the $f(Q)$ model:
\bea
w_{\rm eff}\vert_{z=0}=-0.987_{-0.027}^{+0.032}
\eea
Once again we find an indication that the best fit values of the parameters of  our model are very similar to those of $\Lambda$CDM, in fact, $w_{\rm eff}=-1$ is perfectly inside the $1\sigma$ interval.

\section{Summary and conclusions}\label{Conclusions}

Recent works in the field have been inspired by the realization that, in the  symmetric teleparallel framework, the GR Lagrangian density can be written by basically replacing the scalar of curvature built from the   Levi-Civita connection with the non-metricity $Q$ (up to small details which are not really relevant for a summary level of discussion).

The former framework can be generalized upon replacement of $Q$ with a specific $f(Q)$ given by Eq.~\eqref{fqgr} which reproduces the $\Lambda$CDM
background behavior. Interestingly, the fact that this new setup is not exactly a GR one might have implications at the perturbative level, which is (also) beyond our specific concerns.
The particular action considered at this intermediate test allows us to make progress towards yet another form of  $f(Q)$, which becomes the core of the present work; we present it in Eq.~\eqref{f} and it lets draw an exact expression for the  Hubble parameter under some parameter specifications. The grand picture of this choice is that it resembles standard evolutions  and is therefore worth testing observationally.
To that end we resort to the
MCMC method as to  constraint the free parameters in the theory and compare the values obtained with those of the $\Lambda$CDM scenario.

Our main conclusion is that the parameter which encodes the difference between the two evolutions at the background level is very close to zero when all data sets are combined. This parameter, which we have dubbed $\Omega_Q$, gets positive best fit values for some data sets while it is negative for others, but in all cases the errors make the best fit  perfectly compatible with a null value; thus an overall smaller best fit (almost zero value) is a most admissible consequence. The same conclusion follows from the Bayesian evidence: according to Jeffreys' scale no model is preferred over the other.

A complementary study of the cosmographic parameters yields values which again only reflect the striking similarity of the best fits between $\Lambda$CDM and our $f(Q)$. Note in this respect that, for the modified gravity model, these parameters are more poorly constrained as their complexity penalizes error propagation. Finally, and for the sake of further interpretation of the kind of evolution our best fit scenario depicts, we have computed the corresponding  $w_{\rm eff}$. As expected, the value is very close to $-1$.

For all these reasons we have seen  that a yet another (promising/intriguing) cosmological candidate to become an alternative to $\Lambda$CDM cannot be considered a real  challenger, at least at the background level.

\section*{Acknowledgments}
We are grateful to Jos\'e Beltr\'an Jim\'enez and Marco de Cesare for enlightening discussions. IA and RL thank Fernando D\'\i ez Varela for motivating conversations. This article is based upon work from CANTATA COST (European Cooperation in Science and Technology) action CA15117,  EU Framework Programme Horizon 2020. IA was funded by Fundação para a Ciência e a Tecnologia (FCT) grant number PD/BD/114435/2016 under the IDPASC PhD Program. RL was supported by the Spanish Ministry of Science and Innovation through research projects No. FIS2017-85076-P (comprising FEDER funds) and also by the Basque Govern\-ment through research pro ject No. GIC17/116-IT956- 16.
%
%






%

\end{document}